\documentclass[aos,MSNbibl,nameyear,dvips]{arximspdf}
\usepackage{mathbh}

\doi{10.1214/11-AOS955} 
\volume{40}
\issue{2}
\pubyear{2012}
\firstpage{964}
\lastpage{995}

\makeatletter

\newcommand{\KL}{\mathit{KL}}
\newcommand{\cal}{\mathcal}

\newtheorem{theorem}{Theorem}[section]
\newtheorem{lemma}{Lemma}

\newcommand{\Rplus}{\mathbb{R}^{+}}
\newcommand{\R}{\mathbb{R}}
\newcommand{\I}{\mathbh{1}}

\newcommand{\Xn}{\mathbf{X}_n}
\newcommand{\N}{\mathbb{N}}
\newcommand{\la}{\lambda}
\newcommand{\bla}{{\bolds\lambda}}
\newcommand{\bzero}{{\mathbf{0}}}

\newcommand{\F}{{\cal F}}
\newcommand{\A}{{\cal A}}
\newcommand{\B}{{\cal B}}
\newcommand{\So}{{\cal S}}
\newcommand{\Cont}{{\cal C}^0_+[0,\pi]}
\newcommand{\eqdef}{\stackrel{\Delta}{=}}

\newcommand{\id}{\mathbf{I}_n}
\newcommand{\tr}{\operatorname{tr}}
\newcommand{\bX}{\mathbf{X}}
\newcommand{\equadef}{\stackrel{\Delta}{=}}

\makeatother

\begin{document}
\begin{frontmatter}

\title{Bayesian nonparametric estimation of the spectral density of a long
or intermediate memory Gaussian~process\thanksref{T1}}
\runtitle{Bayesian estimation of a long memory Gaussian process}

\thankstext{T1}{Supported by the
800-2007--2010 Grant ANR-07-BLAN-0237-01 ``SP Bayes.''}

\begin{aug}
\author[A]{\fnms{Judith} \snm{Rousseau}\corref{}\ead[label=e1]{rousseau@ceremade.dauphine.fr}},
\author[A]{\fnms{Nicolas} \snm{Chopin}\ead[label=e2]{nicolas.chopin@ensae.fr}}
\and
\author[B]{\fnms{Brunero} \snm{Liseo}\thanksref{t2}\ead[label=e3]{brunero.liseo@uniroma1.it}}
\runauthor{J. Rousseau, N. Chopin and B. Liseo}
\affiliation{Universit\'{e} Paris Dauphine and CREST-ENSAE, Paris,
CREST-ENSAE, Paris, and~Sapienza Universit\`{a} di Roma}
\address[A]{J. Rousseau\\
N. Chopin\\
CREST-ENSAE\\
3, Avenue Pierre Larousse \\
92245 Malakoff\\
France \\
\printead{e1}\\
\hphantom{E-mail: }\printead*{e2}}
\address[B]{B. Liseo\\
Sapienza Universit\`{a} di Roma\\
Via del Castro Laurenziano \\
9 00161 Roma\\
Italy \\
\printead{e3}} 
\end{aug}

\thankstext{t2}{Supported by the CEREMADE, Universit\'{e} Paris
Dauphine.}

\received{\smonth{7} \syear{2010}}
\revised{\smonth{11} \syear{2011}}

%
\begin{abstract}
A stationary Gaussian process is said to be long-range dependent
(resp.,
anti-persistent) if its spectral density $f(\lambda)$ can be written as
$f(\lambda)=|\lambda|^{-2d}g(|\lambda|)$, where $0<d<1/2$ (resp.,
$-1/2<d<0$), and $g$ is continuous and positive. We propose a novel
Bayesian nonparametric approach for the estimation of the spectral
density of such processes. We prove posterior consistency for both $d$
and $g$, under appropriate conditions on the prior distribution. We
establish the rate of convergence for a general class of priors and
apply our results to the family of fractionally exponential priors. Our
approach is based on the true likelihood and does not resort to
Whittle's approximation.
\end{abstract}

%

\end{frontmatter}

\section{Introduction}

Let $\bX=\{X_{t}, t=1,2,\ldots\}$ be a real-valued stationary zero-mean
Gaussian random process, with spectral density $f$, and covariance
function $\gamma_{f}(\tau)=E(X_{t}X_{t+\tau})$, so that
%
%
\begin{equation}\label{eqgammaf}
\gamma_{f}(\tau)=\int_{-\pi}^{\pi}f(\lambda)e^{i\tau\lambda
}\,d\lambda
\qquad(\tau=0,\pm1,\pm2,\ldots).
\end{equation}
This process is long-range dependent (resp., anti-persistent) if there
exist $C>0$ and a value $d$, $0<d<1/2$ (resp., $-1/2<d<0$), such
that $f(\lambda)|\lambda|^{2d}\rightarrow C$ when $\la\rightarrow0$.
This may be conveniently rewritten as $f(\lambda)=\lambda
^{-2d}g(|\lambda|)$,
where $g\dvtx[0,\pi]\rightarrow\Rplus$ is a continuous positive function.

Interest in long-range dependent and anti-persistent time series has
increased steadily in the last fifteen years; see \citet{ber94}
for a comprehensive introduction and \citet{dou03} for a review
of theoretical aspects and fields of applications, including telecommunications,
economics, finance, astrophysics, medicine and hydrology. Research
in parametric inference for long and intermediate memory processes
have been developed by \citet{mb68}, \citet{mw69}, \citet{ft86},
\citet{dhl89}, \citet{gt99}, \citet{gewe83} and
\citet{ber93},
among others. Unfortunately, parametric inference can be highly biased
under mis-specification of the true model. This limitation makes semiparametric
approaches particularly appealing [\citet{rob95a}].

Under the representation $f(\lambda)=|\lambda|^{-2d}g(|\lambda|)$,
one may like to estimate~$d$ as a measure of long-range dependence,
without resorting to parametric assumptions on the nuisance parameter
$g$.
However, the existing procedures [see the review of \citet{bardet03}]
either exploit the regression structure of the log-spectral density in
a small neighborhood
of the origin [\citet{rob95a}], or use an approximate likelihood function
based on Whittle's approximation [\citet{whi62}], where the original
vector of observations $\Xn=(X_{1},X_{2},\ldots,X_{n})$ gets transformed
into the periodogram $I(\lambda)$ computed at the Fourier frequencies
$\lambda_{j}=2\pi j/n, j=1,2,\ldots,n$, and the artificial observations
$I(\lambda_{1}),\ldots,I(\lambda_{n})$ are, under short range dependence,
approximately independent. Whittle's approximation is very convenient; the
``observations'' $I(\lambda_j)/f(\lambda_j)$ are
approximately independent and identically distributed under
short-range dependence. Unfortunately, this property does not hold under
long-range dependence for the lowest frequencies [\citet{rob95b}].

We propose a Bayesian nonparametric approach to the estimation of
the spectral density of the stationary Gaussian process based on the
true likelihood, without resorting to Whittle's approximation. We
study the asymptotic properties of our procedure, including consistency
and rates of convergence. Our study is based on standard tools for
an asymptotic analysis of Bayesian approaches [e.g., \citet{ggvdv01}];
that is, quantities of interest are the prior probability of a small
neighborhood
around the true spectral density, and some kind of entropy measure
for the prior distribution. Most technical details differ, however,
because of the long-range dependence.

The plan is as follows. In Section~\ref{secnotations},
we introduce the model and the notation. In Section~\ref{secconsG},
we provide a general theorem that states sufficient conditions to
ensure consistency of the posterior distribution, and of several Bayes
estimators. We also introduce the class of FEXP (Fractional Exponential)
priors, based on the FEXP representation of \citet{robi91}, and
show that such prior distributions fulfill these sufficient conditions
for posterior consistency. In Section~\ref{secrates}, we study the
rate of convergence
of the posterior in the general case, and specialize our results for
the FEXP class. Section~\ref{secproofs} gives the proofs of the main
theorems of the two previous sections.
Section~\ref{secDiscussion} discusses further research. The \hyperref
[app]{Appendix}
and the supplement contain technical lemmas.

\section{Model and notation}\label{secnotations}

The model consists of an observed vector $\Xn=(X_{1},\ldots,X_{n})$
of $n$ realizations from a zero-mean Gaussian stationary
process, with spectral density $f$. The likelihood function
is
%
%
\begin{equation}\label{densgaus}
\varphi(\mathbf{X}_{n};f)=(2\pi)^{-n/2}|T_{n}(f)|^{-1/2}\exp\bigl\{
-\tfrac
{1}{2}\Xn^t
T_{n}(f)^{-1}\mathbf{X}_{n}\bigr\},
\end{equation}
where $T_{n}(f)=[\gamma_{f}(j-k)]_{1\leq j,k\leq n}$ is the Toeplitz
matrix associated to $\gamma_{f}$;
see~(\ref{eqgammaf}). This model is parametrized by the pair $(d,g)$,
which defines $f=F(d,g)$
through the factorization
\begin{eqnarray*}
&F\dvtx(-1/2,1/2)\times\Cont \rightarrow \F,&\\
&(d,g) \rightarrow f\dvtx f(\lambda)=|\lambda|^{-2d}g(|\lambda|),&
\end{eqnarray*}
where $\Cont$ is the set of continuous, nonnegative functions over
$[0,\pi]$,
and $\F$ denotes the set of spectral densities, that is, the set
of even functions $f\dvtx[-\pi,\pi]\rightarrow\R^{+}$ such that $\int
_{-\pi
}^{\pi} f(\lambda)\, d\lambda<+\infty$.

The model is completed with a nonparametric prior distribution $\pi$
for $(d,g)\in(-1/2,1/2)\times{\cal C}^{0}_+[0,\pi]$. (There should
be no confusion whether $\pi$ refers to either the number or the
prior distribution in the rest of the paper.) All our results will
assume that the model is valid for some ``true'' parameter
$(d_{0},g_{0})$, associated to some ``true'' spectral density
$f_{0}=F(d_{0},g_{0})$, where $d_{0}\in(-1/2,1/2)$; conditions on
$g_{0}$ are detailed in the next section.

The Kullback--Leibler
divergence for finite $n$ is defined as
\begin{eqnarray*}
\KL_{n}(f_{0};f) & = & \frac{1}{n}\int_{\R^{n}}\varphi(\mathbf{X}_{n};f_{0})
\{ \log\varphi(\mathbf{X}_{n};f_{0})-\log\varphi(\mathbf
{X}_{n};f)\} \,d\Xn\\
& = & \frac{1}{2n}\{ \tr[T_{n}(f_{0})T_{n}^{-1}(f)-\id
]
-\log\det[T_{n}(f_{0})T_{n}^{-1}(f)]\},
\end{eqnarray*}
where $\id$ represents the identity matrix of order $n$.
We also define a symmetrized version of $\KL_{n}$, and its limit as
$n\rightarrow\infty$,
\begin{eqnarray*}
h_{n}(f_{0},f)&=&\KL_{n}(f_{0};f)+\KL_{n}(f;f_{0}),
\\
h(f_{0},f)&=&\frac{1}{4\pi}\int_{-\pi}^{\pi}\biggl[\frac
{f_{0}(\lambda
)}{f(\lambda)}+\frac{f(\lambda)}{f_{0}(\lambda)}-2\biggr]\,d\lambda
= \frac{1}{2\pi} \int_0^{\pi} \biggl( \frac{f_0(\lambda
)}{f(\lambda)}
-1\biggr)^2 \frac{f(\lambda)}{f_0(\lambda)} \,d\lambda.
\end{eqnarray*}
For technical reasons, we also define the pseudo-distance
\[
b_n(f_0,f) = \frac{1}{n} \tr\bigl[ \bigl(
T_n(f)^{-1}T_n(f_0-f)
\bigr)^2 \bigr]
\]
and its limit as $n\rightarrow+\infty$,
\[
b(f_0,f) = \frac{1}{4\pi} \int_{-\pi}^{\pi} \biggl( \frac
{f_0(\lambda
)}{f(\lambda)} -1\biggr)^2\,d\lambda.
\]

Of course, asymptotic pseudo-distances are easier to interpret. In
particular, our consistency results are expressed in terms of the
standard distance~$h$ and posterior concentration results in the
case of FEXP-type priors (see Theorem~\ref{42}) are expressed in terms
of the distance $l(\cdot,\cdot)$ defined in (\ref{defL2}). The
Kullback--Leibler divergence arises naturally in the study of
asymptotic properties of the posterior distribution. The divergence
measure $b_n(\cdot,\cdot)$ is the variance under $f_0$ of
$\log\varphi(\mathbf{X}_{n};f_{0})-\log\varphi(\mathbf{X}_{n};f)$
and is also a common tool in such studies; see, for instance,
\citet{gvdv06}. The symmetrized Kullback--Leibler divergence, $h_n$
is also encountered in Bayesian statistics and is sometimes referred
to as the $J$ divergence; see, for instance, \citet{jeffreys46}.

We also consider the $\mathrm{L}^2$ distance between spectral
log-densities, which is in particular used in \citet{mousou03},
%
%
\begin{equation}\label{defL2}
\ell(f_{0},f) = \int_{-\pi}^{\pi}\{ \log f_{0}(\lambda
)-\log
f(\lambda)\} ^{2}\,d\lambda.
\end{equation}
The advantage of $l$ is that it always exists (for the models
considered here)
whereas the $\mathrm{L}^2$ distance between spectral densities may not.

\section{Consistency}\label{secconsG}

We first state and prove the strong consistency of the posterior distribution
under very general conditions on both $\pi$ and $f_{0}=F(d_{0},g_{0})$;
that is, as $n\rightarrow\infty$, and for $\varepsilon>0$ small enough,
\[
P^{\pi}[\mathcal{A}_{\varepsilon}|\mathbf{X}_{n}]\rightarrow
1\qquad \mbox{a.s.},
\]
where $P^{\pi}[\cdot|\mathbf{X}_{n}]$ denotes posterior probabilities
associated with prior $\pi$, and
\[
\mathcal{A}_{\varepsilon}=\{(d,g)\in(-1/2,1/2)\times\Cont
\dvtx h(f_{0},F(d,g))\leq\varepsilon\}.
\]
From this, we shall deduce the consistency of Bayes estimators of
$f$ and $d$. Finally, we shall introduce the class of FEXP priors,
and show that they allow for posterior consistency.

\subsection{Main result}\label{subsecmain}

Consider the following sets:
\begin{eqnarray*}
{\cal G}(m,M) & = & \{ g\in\mathcal C^0[0,\pi] \dvtx m\leq g\leq
M\}; \\
{\cal G}(m,M,L,\rho) & = & \{ g\in{\cal G}(m,M)\dvtx
|g(\lambda
)-g(\lambda')|\leq L|\lambda-\lambda'|^{\rho}\}; \\
{\cal G}(t,m,M,L,\rho) & = & [-1/2+t,1/2-t]\times{\cal
G}(m,M,L,\rho)
\end{eqnarray*}
for $\rho\in(0,1]$, $L>0$, $m\leq
M$, $t\in(0,1/2)$. Restricting the parameter space to such sets makes
the model identifiable (boundedness of $g$, provided $m>0$), and
ensures that normalized traces of products of Toeplitz matrices that
appear in the distances defined in the previous section converge
(H\"older inequality). We now state our main consistency result.
%
%
\begin{theorem} \label{thcons} For $\varepsilon>0$ small enough,
\[
P^{\pi}[\mathcal{A}_{\varepsilon}|\mathbf{X}_{n}
]\rightarrow
1\qquad \mbox{a.s.}
\]
as $n\rightarrow+\infty$, provided the following conditions are fulfilled:
\begin{enumerate}[(4)]
\item[(1)] There exist $t,m,M,L>0$, $\rho\in(0,1]$, such that 
the set ${\cal G}(t,m,M,L,\rho)$ contains both the pair $(d_{0},g_{0})$
that defines the true spectral density $f_{0}=F(d_{0},g_{0})$ and
the support of the prior distribution $\pi$.
\item[(2)] For all $\varepsilon>0$, $\pi({\cal B}_{\varepsilon
})>0$, where
${\cal B}_{\varepsilon}$ is defined by
\[
{\cal B}_{\varepsilon}=\{ (d,g)\in{\cal G}(t,m,M,L,\rho
)\dvtx h(f_{0},F(d,g))\leq\varepsilon,16|d_{0}-d|<\rho+1-t\} .
\]

\item[(3)] For $\varepsilon>0$ small enough, there exist a sequence
$\F_{n}$
such that $\pi(\F_{n})\geq1-e^{-nr}$, $r>0$, and a net (i.e., a finite
collection)
\[
{\cal H}_{n}\subset\{ (d,g)\in[-1/2+t,1/2-t]\times{\cal
G}(m,M,L,\rho)\dvtx h(f_{0};F(d,g))>\varepsilon/2\}
\]
such that, for $n$ large enough, for all $(d,g)\in\F_{n}\cap
A_{\varepsilon}^{c}$,
$f=F(d,g)$, there exists $(d_{i},g_{i})\in{\cal H}_{n}$,
$f_{i}=F(d_{i},g_{i})$,
such that $8(d_{i}-d)\leq\rho+1-t$, $f\leq f_{i}$, and:

\begin{enumerate}[(a)]
\item[(a)] if $8|d_{i}-d_{0}|\leq\rho+1-t$,
\[
\frac{1}{2\pi}\int_{-\pi}^{\pi}\frac{(f_{i}-f)(\lambda
)}{f_{0}(\lambda
)}\,d\lambda\leq h(f_{0},f_{i})/4;
\]

\item[(b)] if $8(d_{i}-d_{0})>\rho+1-t$,
\[
b(f_{i},f)\leq b(f_{0},f_{i})|{\log\varepsilon}|^{-1};
\]

\item[(c)] otherwise, if $8(d_{0}-d_{i})>\rho+1-t$,
\[
\frac{1}{2\pi}\int_{-\pi}^{\pi}\frac{(f_{i}-f)(\lambda
)}{f_{i}(\lambda
)}\,d\lambda\leq b(f_{i},f_{0})|{\log\varepsilon}|^{-1}.
\]
\end{enumerate}
\item[(4)] The cardinality ${\cal C}_{n}$ of the net ${\cal H}_{n}$
defined above is such that $\log{\cal C}_{n}\leq n\varepsilon
/\log(\varepsilon)$.
\end{enumerate}
\end{theorem}

A proof is given in Section~\ref{appprthcons}. Note that, in the
above definition of
the net $\mathcal{H}_{n}$, the $|{\log\varepsilon}|$ terms are here only
to avoid writing inequalities in terms of awkward constants in the
form $m/M$. If need be, we can replace the $|{\log\varepsilon}|$ by the
correct constants as expressed in Appendix~\ref{apptests}. The
definition of the
above \textit{entropy} is nonstandard. The interest in expressing it in
this general but nonstandard form lies in the difficulty in dealing
with spectral densities which diverge at 0. In practice, the way one
constructs the net $\mathcal H_n$ should vary according to the form of
the prior on the short memory part $g$.

The Bayes estimator associated to loss function $l$ is
\[
\hat{d}=E^{\pi}[d|\mathbf{X}_{n}],\qquad
\hat{g}\dvtx\lambda\rightarrow\exp\{ E^{\pi}[\log
g(\lambda)|\mathbf{X}_{n}]\} ,\qquad
\hat{f}=F(\hat{d},\hat{g}).
\]
Consistency for these point estimates are easily deduced from Theorem
\ref{thcons}, that is, $\hat d\rightarrow d_0$, $l(f_0,\hat
f)\rightarrow0$
a.s. as $n\rightarrow+\infty$; proof of these results are in
the supplementary material [\citet{RouChoLis}, Section 1],
and follow \citet{barron1999consistency}.

\subsection{The FEXP prior}\label{secconsfexp}

Following \citet{hms02}, we consider the FEXP parameterisation of
spectral densities, that is, $f=\tilde{F}(d,k,\theta)$, where
%
%
\begin{eqnarray}\label{eqfexp}
\tilde{F}\dvtx\mathcal{T} & \rightarrow& \F,\nonumber\\[-8pt]\\[-8pt]
(d,k,\theta) & \rightarrow&
f\dvtx f(\lambda)=|1-e^{i\lambda}|^{-2d}\exp\Biggl\{
\sum_{j=0}^{k}\theta_{j}\cos(j\lambda)\Biggr\}
\nonumber
\end{eqnarray}
and $\mathcal{T}=(-1/2+t,1/2-t)\times\{
\bigcup_{k=0}^{+\infty}\{k\}\times\R^{k+1}\} $, for some fixed
$t\in(0,1/2)$. This FEXP representation is equivalent to our previous
representation $f=F(d,g)$, provided $g=\psi^{-d}e^{w}$, $w(\lambda)=
\{ \sum_{j=0}^{k}\theta_{j}\cos(j\lambda) \}$ and
$\psi(\la)=|1-e^{i\lambda}|^{2}/\lambda^{2}=2(1-\cos\la)/\lambda^{2}$
for $\la\neq0$, $\psi(0)=1$. The function $\psi$ is bounded,
infinitely differentiable and positive for $\la\in[0,\pi]$. Thus $g$
and $w$ share the same regularity properties; that is, $w$ is bounded and
H\"older with exponent $\rho$ implies that $g$ is bounded and H\"older with
exponent $\rho$, and vice versa. Under this parameterisation, the prior
distribution $\pi$ is expressed as a trans-dimensional prior
distribution on the random vector $(d,k,\theta)$, which, for
convenience, factorizes as
$\pi_{d}(d)\pi_{k}(k)\pi_{\theta}(\theta|k)$.

We assume that $\pi$ puts mass one on the following Sobolev set:
%
%
\begin{equation}\label{eqsobok}
\So(\beta,L)=\Biggl\{
(d,k,\theta)\in\mathcal{T}\dvtx\sum_{j=0}^{k}\theta
_{j}^{2}(j+1)^{2\beta
}\leq
L\Biggr\}
\end{equation}
for some $\beta>1/2$, $L>0$. This ensures that the Fourier sum
$w$, and thus the short-memory component $g$ of the spectral density
$f$, as explained above, belong to some set $\mathcal{G}(m,M,L',\rho)$,
that is, both $w$ and $g$ are bounded and H\"older, for $\rho<\beta-1/2$.
To see this, note that, for $(d,k,\theta)\in\So(\beta,L)$,
%
%
\begin{eqnarray}\label{eqgeneralinequalitysum}
\sum_{j=0}^{k}|\theta_{j}|j^{r} & \leq& \sum
_{j=0}^{k}\theta
_{j}^{2}(j+1)^{2\beta}+\sum_{j=0}^{k}|\theta_{j}
|j^{r}\I
\bigl(|\theta_{j}|j^{r}\geq\theta_{j}^{2}(j+1)^{2\beta}\bigr)\nonumber
\nonumber\\[-8pt]\\[-8pt]
& \leq& L+\sum_{j=0}^{+\infty}(j+1)^{2r-2\beta}<+\infty,\nonumber
\end{eqnarray}
provided $2r-2\beta<-1$. By taking $r=0$, one sees that $w$ is bounded,
and by taking $r=\rho$, for any $\rho$, $0<\rho<\beta- 1/2$,
one sees that $w$ is H\"older, with coefficient $\rho$, since, for
$\lambda$, $\lambda'\in[-\pi,\pi]$,
\begin{eqnarray*}
|w(\lambda)-w(\lambda')| & \leq& 2\sum
_{j=0}^{k}|\theta
_{j}|\times|\{ \cos(\lambda j)-\cos(\lambda'j)\}
/2|^{\rho}\\
& \leq& 2^{1-\rho}\Biggl(\sum_{j=0}^{k}|\theta_{j}|j^{\rho}
\Biggr)
|\lambda-\lambda'|^{\rho}.
\end{eqnarray*}
Finally, we assume that $\pi$ assigns positive prior probability
to the intersection of $\So(\beta,L)$ with any rectangle set of the
form
$(a_{d},b_{d})\times\{k\}\times\prod_{j=1}^{k}(a_{\theta
_{j}},b_{\theta_{j}})$.

Alternatively, one could assume that the support of $\pi$ is included
in a set of the form $\{(d,k,\theta)\in\mathcal{T}\dvtx\sum
_{j=0}^{k}|\theta_{j}|j^{\rho}\leq L\}$.
However, Sobolev sets are more natural when dealing with rates of convergence
(see Section~\ref{secfexprate}), and are often considered in the
nonparametric literature, so we restrict our attention to these sets.

In the same spirit, we assume that the true spectral density admits
a~FEXP representation associated to an infinite Fourier series,
\[
f_{0}(\lambda)=|1-e^{i\lambda}|^{-2d_{0}}\exp\Biggl\{ \sum
_{j=0}^{+\infty
}\theta_{0j}\cos(j\lambda)\Biggr\},
\]
that is, $f_{0}=F(d_{0},g_{0})$ with $g_{0}=\psi^{-d_{0}}e^{w_{0}}$
and $w_{0}(\la)= \{ \sum_{j=0}^{+\infty}\theta_{0j}\cos
(j\lambda)
\} $. In
addition, we assume that $w_{0}$ satisfies the same type of Sobolev
inequality, namely
%
%
\begin{equation}\label{eqsoboinf}
L_{0}=\sum_{j=0}^{+\infty}\theta_{0j}^{2}(j+1)^{2\beta}<L<+\infty,
\end{equation}
which, as explained above, implies that $g_{0}\in{\cal G}(m,M,L,\rho)$,
for some well-chosen constants $m,M,L,\rho$. Note that it is essential
to have a strict inequality in (\ref{eqsoboinf}), that is, $L_{0}<L$.
%
%
\begin{theorem}\label{thmfexp-conv} Let $\pi$ be a prior distribution
$\pi_{d}(d)\pi_{k}(k)\pi_{\theta}(\theta|k)$ which fulfills the above
conditions, and, in addition, such that $\pi_{k}(k)\leq\exp(-C k
\log k)$
for some $C>0$ and $k$ large enough. Then the conditions of Theorem
\ref{thcons} are fulfilled, and the posterior distribution is consistent.
\end{theorem}
\begin{pf}
Condition (1) of Theorem~\ref{thcons} is a simple consequence of
(\ref{eqsoboinf}) and (\ref{eqsobok}), as explained above. For
condition (2), we noted [see (\ref{eqgeneralinequalitysum})] that
$\sum_{j=0}^{+\infty}\theta_{0j}^{2}(j+1)^{2\beta}\leq L$ implies
that $\sum_{j=0}^{+\infty}|\theta_{0j}|\leq L'<+\infty$. Let $k$
such that $\sum_{j=k+1}^{\infty}|\theta_{0j}|\leq\varepsilon/14$,
$\theta=(\theta_{0},\ldots,\theta_{k})$ such that
$\sum_{j=0}^{k}|\theta_{0j}-\theta_{j}|\leq\varepsilon/14$, $d$ such
that $|d-d_{0}|\leq\varepsilon/7$, and let
$f=\tilde{F}(d,k,\theta)$. Using Lemma~\ref{lemboundh} (see
Appendix~\ref{sectechlem}) one has $h(f,f_{0})\le\varepsilon$.
Note that it is sufficient to prove that $\pi(\B_{\varepsilon})>0$
for $\varepsilon$ small enough; hence we assume that
$\varepsilon/7<(\rho+1-t)/16$. Thus, condition~(2) is
verified as soon as the intersection of $\So(\beta,L)$ and the
rectangle set
\[
[d_{0}-\varepsilon/7,d_{0}+\varepsilon/7]\times\{k\}\times\prod
_{j=1}^{k}[\theta_{0j}-\varepsilon/14k,\theta_{0j}-\varepsilon/14k]
\]
is assigned positive prior probability. Now consider condition (3).
Let $\varepsilon>0$ and take
\[
\F_{n}=\{ (d,k,\theta)\in\So(\beta,L)\dvtx k\leq k_{n}
\} ,
\]
where $k_{n}=\lfloor\alpha n/\log n\rfloor$, for some $\alpha>0$,
so that, for some $r$ depending on $\alpha$,
$\pi(\mathcal{F}_{n}^{c})\leq\pi_k(k>k_n) \leq e^{-nr}$.
Let $f=\tilde{F}(d,k,\theta)$, $f_{i}=(2e)^{c\varepsilon}\tilde
{F}(d_{i},k,\theta_{i})$,
such that $k\leq k_{n}$, $d_{i}-c\varepsilon\leq d\leq d_{i}$, and
$\sum_{j=0}^{k}|\theta_{j}-\theta_{ij}|\leq c\varepsilon$, for some
$c>0$, then
\begin{eqnarray*}
\frac{f(\lambda)}{f_{i}(\lambda)}&=&(2e)^{-c\varepsilon}
[2(1-\cos
\lambda)]^{d_{i}-d}\exp\Biggl\{ \sum_{j=0}^{k}(\theta
_{j}-\theta
_{ij})\cos(j\lambda)\Biggr\} \leq1,
\\
\frac{f(\lambda)}{f_{i}(\lambda)} &\geq&(1-\cos\lambda
)^{c\varepsilon
}2^{-c\varepsilon}e^{-2c\varepsilon}.
\end{eqnarray*}
If $c$ is small enough, $f_{i}-f$ verifies the three
inequalities considered in condition~(3). The number $\mathcal{C}_{n}$
of functions $f_{i}$ necessary to ensure that, for any~$f$ in the
support of $\pi$, at least one of them verify the above inequalities,
can be bounded by, for $n$ large enough, and some well-chosen constant
$C$,
\begin{eqnarray*}
\mathcal{C}_{n}&\leq& k_{n}(Ck_{n}/\varepsilon)^{k_{n}+2} \leq
k_{n}^{3k_{n}}\\
& \leq& \exp\{ 3\alpha n[1+(\log\alpha-\log\log
n
)/\log n]\} \\
& \leq& \exp\{ 6\alpha n\},
\end{eqnarray*}
so condition (4) is satisfied, provided one takes $\alpha=\varepsilon
/6\log\varepsilon$.
\end{pf}

A convenient default choice for $\pi$ is as follows: $\pi_{d}$ is
uniform over $(-1/2+t,1/2-t)$, $\pi_{k}$ is Poisson and
$\pi_{\theta|k}$ has the following structure: the sum
$S=\sum_{j=0}^{k}\theta_{j}^{2}(j+1)^{2\beta}$ has a Gamma
distribution truncated to interval $[0,L]$, independently of $S$, the
vector $(\theta_{0}^{2},\theta_1^2 2^{2\beta}, \ldots,\theta
_{k}^{2}(k+1)^{2\beta})/S$ is
Dirichlet with some coefficients $\alpha_{1,k},\ldots,\alpha_{k,k}$
and the signs of $\theta_{0},\ldots,\theta_{k}$ have equal
probabilities. In particular one may take $\alpha_{j,k}=1$ for all
$j\leq k$, or, if one needs to generate more regular spectral
densities, $\alpha_{j,k}=j^{-\kappa}$, for some fixed or random
$\kappa>0$. Another interesting choice for the prior on $\theta$ is
the following truncated Gaussian process: for each $k$, and each $j
\leq k$,
$\theta_j \sim\mathcal N(0, \tau_0^2( 1 + j)^{-2\beta})$ independently
apart from the constraint, for some fixed, large $L>0$,
\[
\sum_{j=1}^k (1 + j)^{2\beta} \theta_j^2 \leq L.
\]

Note that we can easily restrict ourselves to the important case
$d\geq0$, that is, processes having long or short memory but not intermediate
memory.

\section{Rates of convergence}
\label{secrates}

In this section we first provide a general theorem relating rates
of convergence of the posterior distribution to conditions on the
prior. These conditions are, in essence, similar to the conditions
obtained in the i.i.d. case [e.g., \citet{ggvdv01}]: that is,
a condition on the prior mass of Kullback--Leibler neighborhoods
of the true spectral density, and an entropy condition on the support
of the prior. We then present results specialized to the FEXP
prior case.

\subsection{Main result}

%
%
\begin{theorem}\label{thratespde} Let $(u_{n})$ be a sequence of
positive numbers such that $u_{n}\rightarrow0$, $nu_{n}\rightarrow
+\infty$
and $\bar{{\cal B}}_{n}$ a sequence of balls belonging to ${\cal
G}(t,m,M,L,\rho)$,
and defined as
\[
\bar{{\cal B}}_{n}=\{ (d,g)\dvtx \KL_{n}(f_{0};F(d,g))\leq u_{n}/4,
b_{n}(f_{0},F(d,g))\leq u_{n}, d_0 \leq d\leq d_{0} + \delta\}
\]
for some $\delta,L>0$, $0<m\leq M$, $\rho\in(0,1]$. Let $\pi$
be a prior which satisfies all the conditions of Theorem~\ref{thcons},
and, in addition, such that:
{\renewcommand\thelonglist{(\arabic{longlist})}
\renewcommand\labellonglist{\thelonglist}
\begin{longlist}
\item \label{ratecond1} For $n$ large enough, $\pi(\bar{{\cal
B}}_{n})\geq\exp(-nu_{n}/2)$.
\item \label{ratecond2} There exists $\varepsilon>0$ and a
sequence of
sets $\bar{\F}_{n}\subset\{(d,g)\dvtx h(F(d,g),f_{0})\leq\varepsilon
\}$,
such that, for $n$ large enough,
\[
\pi\bigl(\bar{\F}_{n}^{c}\cap\{(d,g)\dvtx h(F(d,g),f_{0})\leq
\varepsilon\}
\bigr)\leq\exp(-2nu_{n}).
\]

\item \label{ratecond3} There exists a positive sequence
$(\varepsilon_{n})$,
$\varepsilon_{n}^{2}\geq u_{n}$, $\varepsilon_{n}^{2}\rightarrow0$,
$n\varepsilon_{n}^{2} \geq C\log n$, for some $C>0$, satisfying the following
conditions. Let
\[
{\cal V}_{n,l}=\{(d,g)\in\bar{\F}_{n}; \varepsilon_{n}^{2}l\leq
h_{n}(f_{0},F(d,g))\leq\varepsilon_{n}^{2}(l+1)\}
\]
with $l_{0}\leq l\leq l_{n}$, with fixed $l_{0}\geq2$ and
$l_{n}=\lceil
\varepsilon^{2}/\varepsilon_{n}^{2}\rceil-1$.
For each $l=l_{0},\ldots,l_{n}$, there exists a net (i.e., a finite
collection) $\bar{{\cal H}}_{n,l}\subset{\cal V}_{n,l}$, with
cardinality $\bar{{\cal C}}_{n,l}$, such that for all
$f=F(d,g)$, $(d,g)\in{\cal V}_{n,l}$, there exists $f_{i,l} =
F(d_{i,l}, g_{i,l}) \in\bar{{\cal H}}_{n,l}$
such that $f_{i,l} \geq f$ and
\[
0 \leq g_{i,l}(x) - g(x) \leq l \varepsilon_n^2 g_{i,l} /32,\qquad
0 \leq d_{i,l} - d \leq l \varepsilon_n^2 (\log n)^{-1},
\]
where
\[
\log\bar{\mathcal{C}}_{n,l}\leq n\varepsilon_{n}^{2}l^{\alpha
}\qquad
\mbox{with } \alpha<1.
\]
\end{longlist}}

\noindent Then, there exist $C,C'>0$ such that, for $n$ large enough,
%
%
\begin{eqnarray} \label{riskrate}
E_{0}^{n}\bigl[P^{\pi}\bigl(h_{n}(f_{0},F(d,g)) \geq
l_{0}\varepsilon_{n}^{2}|\Xn\bigr)\bigr]
& \leq& Cn^{-3}+2e^{-C'n\varepsilon_{n}^{2}}\nonumber\\[-8pt]\\[-8pt]
& &{} +e^{-nu_{n}/16}.\nonumber
\end{eqnarray}
\end{theorem}

A proof is given in Section~\ref{appproof-theor-rates}.

The conditions given in Theorem~\ref{thratespde} are similar in
spirit to those considered for rates of convergence of the posterior
distribution in the i.i.d. case. The first condition is a condition
on the prior mass of Kullback--Leibler neighborhoods of the true
spectral density, the second one is necessary to allow for sets with
infinite entropy (some kind of noncompactness) and the third one is
an entropy condition. The inequality (\ref{riskrate}) obtained in
Theorem~\ref{thratespde} is nonasymptotic, in the sense that it
is valid for all $n$. However, the distances considered in Theorem
\ref{thratespde} heavily depend on $n$ and, although they express
the impact of the differences between $f$ and $f_{0}$ on the
observations, they can be difficult to work with. Note that the
metric $h_n$, which is a symmetrized version of the Kullback--Leibler
divergence $\KL_n$, leads to a strong convergence result\vadjust{\goodbreak} since it
implies in particular a similar posterior concentration rate for any
metric smaller than $h_n$, which includes $\KL_n$. For these reasons,
the entropy condition is awkward and cannot be directly transformed
into some more common entropy conditions. To state a result
involving distances between spectral densities that might be more
useful, we need to consider more specific class of priors. In the
next section, we obtain rates of convergence in terms of the $\ell$
distance for the class of FEXP priors introduced in Section
\ref{secconsfexp}. The rates obtained are the optimal rates up to
a $(\log n)$ term, at least on certain classes of spectral
densities. It is to be noted that the calculations used when working
on these classes of priors are actually more involved than those
used to prove Theorem~\ref{thratespde}. This is quite usual when
dealing with rates of convergence of posterior distributions;
however, this is emphasized here by the fact that distances involved
in Theorem~\ref{thratespde} are strongly dependent on $n$. The
method used in the case of the FEXP prior can be extended to other
types of priors.

\subsection{Rates of convergence for the FEXP prior}
\label{secfexprate}

We apply Theorem~\ref{thratespde} to the class of FEXP priors introduced
in Section~\ref{secconsfexp}. Recall that under such a prior a
spectral density $f$ is parametrized as $f=\tilde{F}(d,k,\theta)$;
see (\ref{eqfexp}). We make the same assumptions as in Section
\ref{secconsfexp}.
In particular, the prior $\pi(d,k,\theta)$ factorizes as $\pi
_{d}(d)\pi
_{k}(k)\pi_{\theta}(\theta|k)$;
the right tail of $\pi_{k}$ is such that
\[
\exp\{ -C k\log k \} \leq\pi_{k}(k)\leq\exp\{-C'k\log
k\}
\]
for some $C$, $C'>0$, and for $k$ large enough; and there exists
$\beta>1/2$ such that the Sobolev set $S(\beta,L)$ contains the
support of $\pi$. The last condition means that
$S=\sum_{j=0}^{k}\theta_{j}^{2}(j+1)^{2\beta}\in[0,L]$ with prior
probability one. In addition, we assume that the support of $\pi_{d}$
is $[-1/2+t,1/2-t]$, and, for $d\in[-1/2+t,1/2-t]$, $\pi_{d}(d)\geq
c_{d}>0$. Similarly, we assume that $\pi_{\theta|k}$ is such that the
random variable $S=\sum_{j=0}^{k}\theta_{j}^{2}(j+1)^{2\beta}$ is
independent of $k$, and admits a~probability density $\pi_{S}(s)$
with support $[0,L]$, and such that $\pi_{S}(s)\geq c_{s}>0$ for
$s\in[0,L]$.
%
%
\begin{theorem}\label{42} For the FEXP prior described above,
there exist \mbox{$C,C'>0$} such that, for $n$ large enough,
%
%
\begin{equation}\label{postrate}
E_{0}^{n}\biggl\{ P^{\pi}\biggl[\ell(f,f_{0})>\frac{C\log{n}
}{n^{2\beta/(2\beta+1)}}\Big|\mathbf{X}_{n}\biggr]\biggr\} \leq
\frac
{C}{n^{2}},
\end{equation}
where $f=\tilde{F}(d,k,\theta)$ and
%
%
\begin{equation}\label{riskrate2}
E_{0}^{n}[\ell(\hat{f},f_{0})]\leq\frac{C'(\log
n)}{n^{2\beta
/(2\beta+1)}},
\end{equation}
where $\log{\hat{f}}(\lambda)=E^{\pi}[\log{f}(\lambda
)|\mathbf
{X}_{n}]$.
\end{theorem}

A proof is given in Appendix~\ref{secproof-theor-rates}.\vadjust{\goodbreak}

\section{\texorpdfstring{Proofs of Theorems \protect\ref{thcons} and \protect\ref{thratespde}}
{Proofs of Theorems 3.1 and 4.1}}
\label{secproofs}

\subsection{\texorpdfstring{Proof of Theorem \protect\ref{thcons}}{Proof of Theorem 3.1}}
\label{appprthcons}

For the sake of conciseness, we introduce
the following notation: for any pair $(f,f_0)$ of spectral densities,
\begin{eqnarray*}
A(f_{0},f) &=& T_{n}(f)^{-1}T_{n}(f_{0}),\\
B(f_{0},f) &=&
T_{n}(f_{0})^{1/2}[T_{n}(f)^{-1}-T_{n}(f_{0})^{-1}]T_{n}(f_{0})^{1/2}.
\end{eqnarray*}

The proof borrows ideas from \citet{ggvdv01}. The main difficulty
is to formulate constraints on quantities such as $h_{n}(f,f_{0})$
or $\KL_{n}(f,f_{0})$ in terms of distances between $f,f_{0}$, independent
on $n$, and uniformly over $f$. One has
\[
P^{\pi}[{\A}_{\varepsilon}^{c}|\mathbf{X}_{n}] =
\frac
{\int\I_{\A_{\varepsilon}^{c}}(f)\varphi(\mathbf
{X}_{n};f)/\varphi
(\mathbf{X}_{n};f_{0})\,d\pi(f)}{\int\varphi(\mathbf
{X}_{n};f)/\varphi
(\mathbf{X}_{n};f_{0})\,d\pi(f)}\eqdef\frac{N_{n}}{D_{n}}.
\]
Let $\delta\in(0,\varepsilon)$ and $P_{0}^{n}$ be a generic notation
for probabilities associated to the distribution of $\Xn$, under
the true spectral density $f_{0}=F(d_{0},g_{0})$. One has
%
%
\begin{equation}\label{eqp1pp2}
P_{0}^{n}\{P^{\pi}[\A_{\varepsilon}^{c}|\mathbf
{X}_{n}
]\geq
e^{-n\delta}\} \leq P_{0}^{n}[D_{n}\leq
e^{-n\delta}]+P_{0}^{n}[N_{n}\geq
e^{-2n\delta}],
\end{equation}
so that Theorem~\ref{thcons} follows from bounds on both terms of the
right-hand side of the above inequality.
The following lemma bounds the first term.
%
%
\begin{lemma}
\label{lemDn}
There exists $C>0$ such that
%
%
\begin{equation}\label{eqp1}
P_{0}^{n}[D_{n}\leq e^{-n\delta}]\leq Cn^{-3}.
\end{equation}
\end{lemma}
\begin{pf}
Lemma~\ref{lemunifTR2} implies that, when $n$ is large enough,
$\tilde{{\cal B}}_{n}\supset{\cal B}_{\delta/8}$, where
\[
\tilde{{\cal B}}_{n}=\{(d,g)\in[-1/2+t,1/2-t]\times{\cal
G}(m,M,L,\rho
)\dvtx \KL_{n}(f_{0},F(d,g))\leq\delta/4\},
\]
and condition (2) implies that, for $n$ large enough, $\pi(\tilde
{{\cal B}}_{n})\geq\pi(\mathcal{B}_{\delta/8})\geq2e^{-n\delta/2}$.
Consider the indicator function
\[
\Omega_{n}=\I[-\Xn^{t}\{
T_{n}(f)^{-1}-T_{n}(f_{0})^{-1}
\} \Xn+\log\det A(f_{0},f)>-n\delta]
\]
with implicit arguments $(f,\Xn)$, then, following \citet{ggvdv01},
\begin{eqnarray*}
P_{0}^{n}[D_{n}\leq e^{-n\delta}] & \leq& P_{0}^{n}
\biggl(\int\Omega_{n}\I_{\tilde{{\cal B}}_{n}}(f)\frac{\varphi(\mathbf
{X}_{n};f)}{\varphi(\mathbf{X}_{n};f_{0})}\,d\pi(f)\leq e^{-n\delta
/2}\frac{\pi(\tilde{{\cal B}}_{n})}{2}\biggr)\\
& \leq& P_{0}^{n}\bigl(E^{\pi}\{ \Omega_{n}\I_{\tilde{{\cal
B}}_{n}}(f)\} \leq\pi(\tilde{{\cal
B}}_{n})/2\bigr)\\
& \leq& P_{0}^{n}\bigl(E^{\pi}\{ (1-\Omega_{n})\I_{\tilde
{{\cal
B}}_{n}}(f)\} \geq\pi(\tilde{{\cal
B}}_{n})/2\bigr)\\
& \leq& \frac{2}{\pi(\tilde{{\cal B}}_{n})}\int_{\tilde{{\cal
B}}_{n}}E_{0}^{n}\{ 1-\Omega_{n}\} \,d\pi(f)
\end{eqnarray*}
by Markov's inequality. Besides,
\begin{eqnarray*}
E_{0}^{n}\{ 1-\Omega_{n}\} & = &
P_{0}^{n}\bigl\{\mathbf{X}_{n}^{t}\{
T_{n}(f)^{-1}-T_{n}(f_{0})^{-1}\} \mathbf{X}_{n}-\log\det
A(f_{0},f)>n\delta\bigr\}\\
& = & P_{\mathbf{Y}}\{
\mathbf{Y}^{t}B(f_{0},f)\mathbf{Y}-\tr[B(f_{0},f)
]>D(f_{0},f)\},
\end{eqnarray*}
where $\mathbf{Y}\sim N_{n}(\bzero_{n},\id)$, and, for $f\in\tilde
{\B}_n$,
\[
D(f_{0},f) \equadef n\delta+\log\det A(f_{0},f)-\tr
[B(f_{0},f)]>n\delta/2
\]
thus
\begin{eqnarray*}
E_{0}^{n}[1-\Omega_{n}] & \leq& P_{\mathbf{Y}}\{ \mathbf
{Y}^{t}B(f_{0},f)\mathbf{Y}-\tr[B(f_{0},f)]>n\delta/2\} \\
& \leq& \frac{16}{n^{4}\delta^{4}}E_{\mathbf{Y}}\bigl[\{
\mathbf
{Y}^{t}B(f_{0},f)\mathbf{Y}-\tr[B(f_{0},f)]\} ^{4}\bigr]\\
& \leq& \frac{C}{n^{3}\delta^{4}},
\end{eqnarray*}
which concludes the proof.
\end{pf}

A bound for the second term in (\ref{eqp1pp2}) is obtained
as follows:
%
%
\begin{eqnarray} \label{eqp2p3}
P_{0}^{n}[N_{n}\geq e^{-2n\delta}] & \leq& 2e^{2n\delta
}\pi
(\F_{n}^{c})+p\nonumber\\[-8pt]\\[-8pt]
& \leq& 2e^{-n(r-2\delta)}+p\nonumber
\end{eqnarray}
using condition (3), where
\[
p\eqdef
P_{0}^{n}\biggl[\int\I(A_{\varepsilon}^{c}\cap\F_{n})\frac
{\varphi
(\mathbf{X}_{n};f)}{\varphi(\mathbf{X}_{n};f_{0})}\,d\pi(f)\geq
e^{-2n\delta}/2\biggr].
\]

Assuming $2\delta<r$, we consider the following likelihood ratio
tests for each \mbox{$f_{i}\in{\cal H}_{n}$}, and for some arbitrary values
$\rho_{i}$,
\[
\phi_{i}=\I\{ \Xn^{t}
[T_{n}^{-1}(f_{0})-T_{n}^{-1}(f_{i})]\Xn\geq n\rho_{i}
\} .
\]

Lemmas~\ref{7},~\ref{8} and~\ref{9} given
in Appendix~\ref{apptests} prove that, for each of the three cases in
condition (3) of Theorem~\ref{thcons}, and well-chosen values of
$\rho_{i}$, one has
%
%
\begin{equation}\label{eqEphi}
E_{0}^{n}[\phi_{i}]\leq e^{-nC_{1}\varepsilon},\qquad E_{f}^{n}[1-\phi
_{i}]\leq e^{-nC_{1}\varepsilon}
\end{equation}
for all $f_{i}$, for $f$ close to $f_{i}$ [in the sense defined
in cases (a), (b) and (c) in condition~(3)], where $C_{1}>0$ is a constant
that does not depend on $f_{i}$, and $E_{f}^{n}$ stands\vspace*{1pt} for the
expectation with respect to the likelihood $\varphi(\Xn;f)$. Then
one concludes easily as follows. Let $\phi^{(n)}=\max_{i}\phi_{i}$;
then, using Markov inequality, for $n$ large enough,
%
%
\begin{eqnarray}\label{eqp3}
p & \leq& E_{0}^{n}\bigl[\phi^{(n)}\bigr]+2e^{2n\delta}\int
_{A_{\varepsilon}^{c}\cap\F_{n}}E_{f}\bigl[1-\phi^{(n)}
\bigr]\,d\pi
(f)\nonumber\\[-8pt]\\[-8pt]
& \leq& \mathcal{C}_{n}e^{-nC_{1}\varepsilon}+2e^{2n\delta
-nC_{1}\varepsilon}
\leq e^{-nC_{1}\varepsilon/2},\nonumber
\end{eqnarray}
provided $\delta<C_{1}\varepsilon/4$. Combining (\ref{eqp1}),
(\ref{eqp2p3}) and (\ref{eqp3}), there exists $\delta>0$ such
that
\[
P_{0}^{n}\bigl[P^{\pi}[A_{\varepsilon}^{c}|\mathbf
{X}_{n}]>e^{-n\delta
}\bigr]\leq Cn^{-3}
\]
for $n$ large enough, which implies that $P^{\pi}[A_{\varepsilon
}^{c}|\mathbf{X}_{n}]\rightarrow0$
a.s.

\subsection{\texorpdfstring{Proof of Theorem \protect\ref{thratespde}}{Proof of Theorem 4.1}}
\label{appproof-theor-rates}

This proof uses the same notation as the previous section: $C$,
$C'$ denote generic constants, $f$, $d\pi(f)$ are short-hands for
$f=F(d,g)$, $d\pi(d,g)$, respectively, $A(f,f_0)$ and $B(f,f_0)$ have
the same definition, and so on.
In the proof of Theorem~\ref{thcons}, we showed that
$E_{0}^{n}[P^{\pi}(h(f,f_{0})\geq\varepsilon|\Xn)]\leq
Cn^{-3}$ for $\varepsilon$ small enough, $n$ large enough. Thanks to
the uniform convergence in Lemmas~\ref{lemunifTR1} and~\ref{lemunifTR2}
in Appendix~\ref{sectechn-lemm-contr}, one sees that the same
inequality holds if $h$ is replaced by $h_{n}$. Therefore, to obtain
inequality (\ref{riskrate}), it is sufficient to bound the
expectation of the sum of the following probabilities:
\[
P^{\pi}\bigl((d,g)\in\mathcal{W}_{n,l}|\Xn\bigr)
=\frac{\int\I_{\mathcal{W}_{n,l}}(d,g)({\varphi(\mathbf
{X}_{n};f)}/{\varphi(\mathbf{X}_{n};f_{0})})\,d\pi(f)}
{\int({\varphi(\mathbf{X}_{n};f)}/{\varphi(\mathbf
{X}_{n};f_{0})})\,d\pi
(f)}=\frac{N_{n,l}}{D_{n}}
\]
for $l_{0}\leq l\leq l_{n}$, where ${\cal V}_{n,l}={\cal W}_{n,l}\cap
\bar{{\cal F}}_{n}$ and
\[
{\cal W}_{n,l}=\{ (d,g)\dvtx h(f,f_{0})\leq\varepsilon,
\varepsilon
_{n}^{2}l\leq h_{n}(f_{0},f)\leq\varepsilon_{n}^{2}(l+1)\}.
\]

To prove the theorem one can follow the same lines as in Section \ref
{appprthcons} to show that
%
%
\begin{eqnarray}\label{eqDnEn}\quad
E_{0}^{n}\Biggl[\sum_{l=l_{0}}^{l_{n}}\frac{N_{n,l}}{D_{n}}\Biggr]
& \leq&
P_{0}^{n}(D_{n}\leq
e^{-nu_{n}}/2) \nonumber\\
&&{}+E_{0}^{n}\Biggl[\sum_{l=l_{0}}^{l_{n}}\frac
{N_{n,l}}{D_{n}}\I(D_{n}\geq
e^{-nu_{n}}/2)\Biggr] \\
:\!&=& A_n + B_n.
\nonumber
\end{eqnarray}
Now we show that both $A_n$ and $B_n$ can be bounded.

\subsubsection{Boundedness of $A_n$}

$A_n$ can be bounded as in Lemma~\ref{lemDn}; see
Section~\ref{appprthcons}: in fact,
\begin{eqnarray*}
P_{0}^{n}(D_{n}\leq e^{-nu_{n}}/2)
& \leq& P_{0}^{n}\biggl(D_{n}\leq\frac{e^{-nu_{n}/2}\pi(\bar{{\cal
B}}_{n})}{2}\biggr)\\
& \leq&
\frac{2\int_{\mathcal{B}_{n}}E_{0}^{n}[(1-\Omega
_{n}(f))
]\,d\pi(f)}{\pi(\bar{{\cal B}}_{n})},
\end{eqnarray*}
where $\Omega_{n}$ is the indicator function of
\[
\bigl\{
(\Xn,f);\Xn^{t}\bigl(T_{n}^{-1}(f)-T_{n}^{-1}(f_{0})\bigr)\Xn-\log\det
[A(f_{0},f)]\leq
nu_{n}\bigr\}.
\]
Also note that, for $f\in\bar{{\cal B}}_{n}$,
there exists $s_0>0$ such that for all
$s\leq s_0$,
\[
\id(1 + 2s) - 2 s T_n(f_0)^{1/2}T_n(f)^{-1}T_n(f_0)^{1/2} \geq
\id/2.
\]
Using Chernoff-type inequalities as in Lemma~\ref{7}, one can show that
for $f = F(d$, $g)$, $d \geq d_0$, $g>0$, and for all $0<s\leq s_0$,
\begin{eqnarray*}
E_{0}^{n}[1 - \Omega_{n}]
& \leq& \exp\biggl\{ - snu_n - {s \log}| T_n(f_0) T_n(f)^{-1}| \\
& &\hphantom{\exp\biggl\{}{} - {\frac{1}{2} \log}| \id(1 + 2s) - 2s
T_n(f_0)^{1/2}T_n(f)^{-1}T_n(f_0)^{1/2} | \biggr\} \\
& \leq& \exp\{ - s n u_n + 2s n\KL_n(f_0,f) + 4 s^2n b_n(f_0,f)
\} \\
& \leq&
\exp\biggl\{ - \frac{ s n u_n }{ 2} (1 - 8 s) \biggr\}
\leq e^{-Cn u_n}.
\end{eqnarray*}
In the above derivation, the second inequality comes from a Taylor
expansion in $s$ of
$\log| \id+ 2s (\id- T_n(f_0)^{1/2}T_n(f)^{-1}T_n(f_0)^{1/2} )| $,
the third comes from the definition of $\bar{{\cal B}}_{n}$ and the last
from choosing $s = \min(s_0, 1/16)$. Note that $s_0 \geq m/(M\pi)$
and that the constant $C$ in the above inequality can be chosen as $C
= m / (32 M \pi) $.

\subsubsection{Boundedness of $B_n$}

$B_n$ can be written as
%
%
\begin{eqnarray}\label{eqKn}
B_n &=& E_{0}^{n}\Biggl[\sum_{l=l_{0}}^{l_{n}}\frac
{N_{n,l}}{D_{n}}\I(D_{n}\geq e^{-nu_{n}}/2)(\bar{\phi
}_{l}+1-\bar{\phi}_{l})\Biggr]\nonumber\\[-8pt]\\[-8pt]
& \leq& \sum_{l=l_{0}}^{l_{n}}E_{0}^{n}[\bar{\phi}_{l}
]+2e^{nu_{n}}\sum_{l=l_{0}}^{l_{n}}E_{0}^{n}[N_{n,l}(1-\bar
{\phi
}_{l})],\nonumber
\end{eqnarray}
where $\bar{\phi}_{l}=\max_{i\dvtx f_{i,l}\in\bar{{\cal H}}_{n,l}}\phi
_{i,l}$, $\phi_{i,l}$ is a test function defined
as in Section~\ref{appprthcons},
\begin{eqnarray*}
\phi_{i,l}&=&\I\bigl\{
\mathbf{X}_{n}'\bigl(T_{n}^{-1}(f_{0})-T_{n}^{-1}(f_{i,l})\bigr)\mathbf{X}_{n}
\geq
\tr[\id-T_{n}(f_{0})T_{n}^{-1}(f_{i,l})
]\\
&&\hspace*{179pt}{}+nh_{n}(f_{0},f_{i,l})/4\bigr\}.
\end{eqnarray*}
We now show that both terms in the right-hand side of (\ref{eqKn})
are bounded.
For the first term, we first derive a bound for the logarithm of
$E_{0}^{n}[\phi_{i,l}] $:
using inequality (\ref{eqE0phi}) in Lemma~\ref{7}, one has
%
%
\begin{equation} \label{test1}
\log E_{0}^{n}[\phi_{i,l}] \leq- C n h_n(f_0,f_i) \min\biggl( \frac{
h_n(f_0,f_i) }{ b_n(f_0, f_i) } , 1\biggr)
\end{equation}
for some universal constant $C$, and $n$ large enough. In addition, one has
\[
\frac{ b_n(f_0,f_i) }{ h_n(f_0, f_i) }
\leq\|T_n(f_0)^{1/2}T_n(f_i)^{-1/2} \|^2
\leq C' n^{2\max( d_0- d_i, 0)}.
\]
The first inequality comes from Lemma~\ref{lemhnbn} of
Appendix~\ref{secmatrixineq}, and
the second inequality comes from Lemma 3 in\vadjust{\goodbreak}
\citet{LRR09}. Hence for all $C>0$, if
$2|d_0 - d_i| \leq C /\log n$, $b_n(f_0,f_i) \leq C'e^C h_n(f_0,f_i)$.
Moreover for all $\delta>0$,
there exists $C_\delta>0$ such that if $2|d_0 - d_i| > C_\delta
(\log n )^{-1}$, then $h_n(f_0,f_i) \geq n^{-\delta}$. Indeed,
equation (\ref{eqwhit2}) of Lemma~\ref{whittles} implies that if
$h_n(f_0,f_i)\geq
\varepsilon_n^2$, then
\[
h_n(f_0,f_i) \geq\frac{C}{n} \tr[
T_n(f_0^{-1})T_n(f_i-f_0)T_n(f_i^{-1}) T_n(f_i-f_0) ]
\]
and Lemma~\ref{lemLPunif} (see Appendix~\ref{secALP}) implies that,
for all $a>0$,
\begin{eqnarray*}
&&\biggl| \frac{1}{n} \tr[ T_n(f_0^{-1})T_n(f_i-f_0)T_n(f_i^{-1})
T_n(f_i-f_0) ] - (2\pi)^{3} \int_{-\pi}^\pi\frac{
(f_i-f_0)^2 }{ f_i f_0} \,d\lambda\biggr|\\
&&\qquad \leq n^{-\rho+a}.
\end{eqnarray*}

Lemma~\ref{lemdgebge} in Appendix~\ref{sectechlem} implies that there
exists $a>0$ such that, if
$2|d_0 - d_i| > C_\delta(\log n )^{-1}$,
\[
\int_{-\pi}^\pi\frac{ (f_i-f_0)^2 }{ f_i f_0} \,dx \geq C e^{- a \log n
/ C_\delta} \geq n^{-\delta}
\]
as soon as $C_\delta$ is large enough. Choosing $\delta< \rho$ we
finally obtain that
$h_n(f_0,\allowbreak f_i) \geq C' n^{-\delta}$.
This and the definition of $\bar{\mathcal H}_{n,l}$ implies that $l
\geq C'n^{-\delta} \varepsilon_n^{-2}$, and therefore
$ l
n^{-\max(d_0-d_i,0)} \geq2l^\alpha/C'$,
for all $\alpha< 1$ as soon as $|d_0 - d_i| $ is small enough.
This implies that (\ref{test1}) becomes
\[
\log E_{0}^{n}[\phi_{i,l}] \leq- c l \varepsilon_n^2
n^{1-\max(d_0-d_i,0)}
\leq-2 n\varepsilon_n^2 l^\alpha.
\]
Also, condition~\ref{ratecond3} implies that
\[
E_{0}^{n}[\bar{\phi}_{l}]
\leq\sum_{i}E_{0}^{n}[\phi_{i,l}]\leq\bar{{\cal
C}}_{n,l}\exp
\{-2n\varepsilon_{n}^{2}l^\alpha\}\leq\exp\{- n\varepsilon
_{n}^{2}l^\alpha\}
\]
so that $\sum_{l}E_{0}^{n}[\bar{\phi}_{l}]\leq2 \exp\{-
n\varepsilon_{n}^{2}l_{0}^\alpha\}$
for $n$ large enough.

The second term of the right-hand side of (\ref{eqKn}) is bounded by
considering that,
from condition~\ref{ratecond3} on $f$ and $f_{i,l}$, one has
\[
0\leq f_{i,l} - f \leq h_n(f_0, f_{i,l}) f_{i,l} \biggl( \frac{
\pi^{2(d_i-d)}}{32} + \frac{ 2 | {{\log}|\lambda|} | }{\log n} \biggr)
\]
for $n$ is large enough; hence
$\tr A(f_{i,l}-f,f_0) \leq n h_n(f_0,f_{i,l})/4$, and
we obtain the first part of equation (\ref{secondtypecasea}),
\[
\log E_{f}^{n}[1-\phi_{i,l}]
\leq-\frac{n}{64}\min\biggl( \frac
{h_{n}(f_{0},f_{i,l})^{2}}{b_{n}(f,f_{0})},4h_{n}(f_{0},f_{i,l})\biggr).
\]
We also have
\[
b_n(f,f_0) \leq b_n(f_{i,l},f_0) + \frac{ h_n^2(f_{i,l},f_0) }{ 32} + 2
\sqrt{b_n(f_0,f_{i,l}) } h_n(f_{i,l},f_0),
\]
hence $ \log E_{f}^{n}[1-\phi_{i,l}] \leq- c n
l^\alpha\varepsilon_n^2 $, using the same arguments as before, and
\begin{eqnarray*}
\sum_{l=l_{0}}^{l_{n}}{E_{0}^{n}[(1-\bar{\phi
}_{l})N_{n,l}]}
&=& \int\Biggl\{ \sum_{l=l_{0}}^{l_{n}}\I_{{\cal
W}_{n,l}}(f)E_{f}(1-\bar
{\phi}_{l})\Biggr\} \, d\pi(f)\\
&\leq& P^{\pi}\bigl(f\in\mathcal{F}_{n}^{c}\cap\{
h(f,f_{0})\leq
\varepsilon\} \bigr)\\
&&{}
+\sum_{l=l_{0}}^{l_{n}}\int\I_{{\cal V}_{n,l}}(f)E_{f}^{n}(1-\bar
{\phi
}_{l})\, d\pi(f)\\
&\leq& e^{-n\varepsilon_{n}^{2}}+\sum
_{l=l_{0}}^{l_{n}}e^{-Cn\varepsilon
_{n}^{2}l^\alpha}\leq2e^{-n\varepsilon_{n}^{2}}.
\end{eqnarray*}

\section{Discussion}\label{secDiscussion}

In this paper we have considered the theoretical properties of
Bayesian nonparametric estimates of the spectral density for
Gaussian long memory processes. Some general conditions on the prior
and on the true spectral density are provided to ensure consistency
and to determine concentration rates of the posterior distributions
in terms of the pseudo-metric $h_n(f_0,f)$. To derive a posterior
concentration rate in terms of a more common metric such as
$l(\cdot,\cdot)$, we have considered a specific family of priors based of
the FEXP models that are also used in the frequentist literature.
Gaussian long memory processes lead to complex behaviors, which
makes the derivation of concentration rates a difficult task. This
paper is thus a~step in the direction of better understanding the
asymptotic behavior of the posterior distribution in such models
and could be applied to various types of priors on the short memory
part---other than the FEXP priors.

The rates we have derived are optimal (up to a $\log n$ term) in
Sobolev balls but not adaptive since the estimation procedure
depends on the smoothness~$\beta$. Another limitation is that the
prior is restricted to Sobolev balls with fixed though large radius.
But, even in the parametric framework, current asymptotic results on
likelihood-based approaches all assume the parameter space to be
compact. The technical reason is that all these results rely on the
short memory part of the spectral density being uniformly bounded.

A related and fundamental problem is the practical implementation of
the model described in the paper. \citet{lr06} adopted a~Population
MC algorithm which easily deals with the trans-dimensional parameter
space issue. We are currently working on alternative computational
approaches.


%
\begin{appendix}\label{app}

\section{Technical lemmas on convergence rates of~products of Toeplitz
matrices}
\label{sectechn-lemm-contr}

We first give a set of inequalities on norms of matrices that are
useful throughout the proofs. We then give three technical lemmas on
the uniform convergence of traces of products of Toeplitz matrices, in
the spirit of \citet{jud03} and \citet{LRR09}, but extending those
previous results to functional classes instead of parametric classes.

\subsection{Some matrix inequalities}
\label{secmatrixineq}

Let $A$ and $B$ be $n$-dimensional matrices. We consider the following
two norms:
\[
|A|^2 = \tr[ A A^t ] , \qquad\| A\|^2 = \sup_{ |x|=1}
(
x^t A A^t x).
\]
We recall that:
$|{\tr}[ A B] | \leq|A| |B|$, $|A B| \leq\| A\| |B|$,
$|A| \leq\| A\|$, $\|A B\| \leq\| A\| \|B\|$.
Using these inequalities we prove the following basic lemma:
%
%
\begin{lemma} \label{lemhnbn}
Let $f_1, f_2$ be two spectral densities, then
\[
2 nb_n(f_1,f_2) \leq n \| T_n(f_2)^{-1/2} T_n(f_1)^{1/2} \|^2 h_n(f_1,f_2).
\]
\end{lemma}
\begin{pf}
One has
\begin{eqnarray*}
&&2 nb_n(f_1,f_2) \\
&&\qquad= \tr\bigl[ T_n(f_1)^{1/2}T_n(f_2)^{-1} T_n(f_1)^{1/2} \bigl(
T_n(f_1)^{-1/2}T_n(f_1 - f_2) T_n(f_2)^{-1/2} \bigr)^2 \bigr] \\
&&\qquad= \bigl|T_n(f_2)^{-1/2} T_n(f_1)^{1/2} \bigl( T_n(f_1)^{-1/2}T_n(f_1
- f_2) T_n(f_2)^{-1/2} \bigr)\bigr|^2 \\
&&\qquad\leq \|T_n(f_2)^{-1/2} T_n(f_1)^{1/2}\|^2 |
T_n(f_2)^{-1/2}T_n(f_1 - f_2) T_n(f_2)^{-1/2} |^2 \\
&&\qquad= n \| T_n(f_2)^{-1/2} T_n(f_1)^{1/2} \|^2 h_n(f_1,f_2).
\end{eqnarray*}
\upqed
\end{pf}

\subsection{\texorpdfstring{Uniform convergence: Lemmas \protect\ref{lemunifTR1} and
\protect\ref{lemunifTR2}}
{Uniform convergence: Lemmas 3 and 4}}

We state two technical lemmas, which are extensions of \citet{jud03}
on uniform convergence of traces of Toeplitz matrices, and which are
repeatedly used in the paper.
%
%
\begin{lemma} \label{lemunifTR1}
Let $t>0$, $M,L>0$ and $\rho\in(0,1]$, let $p$ be a positive integer,
we have, as $n\rightarrow+\infty$,
\begin{eqnarray*}
\mathop{\mathop{\mathop{\sup_{f_{i}=F(d_{1},g_{i}),
f_{i}'=F(d_{2},g_{i}')}}_{2p(d_{1}+d_{2})\leq1-t}}_{
g_{i}\in\mathcal{G}(-M,M,L,\rho)}}_{
g_{i}'\in\mathcal{G}(-M,M,L,\rho)}
\Biggl|\frac{1}{n}\tr\Biggl[\prod
_{i=1}^{p}T_{n}(f_{i})T_{n}(f_{i}')\Biggr]
- \frac{ \int_{-\pi}^{\pi}\prod_{i=1}^{p}f_{i}(\lambda
)f_{i}'(\lambda
) \, d\lambda}{ (2 \pi)^{1 - 2p}} \Biggr|\rightarrow0.
\end{eqnarray*}
\end{lemma}

This lemma is a direct adaptation from \citet{jud03};
the only nonobvious part is the change from the condition of continuous
differentiability in that paper to the Lipschitz condition of order
$\rho$. This different assumption
affects only equation (30) of \citet{jud03}, with $\eta_{n}$ replaced
by $\eta_{n}^{\rho}$, which does not change the convergence
results.\vadjust{\goodbreak}
%
%
\begin{lemma}\label{lemunifTR2} Let $t>0$, $M,L,m>0$ and $\rho
_{1},\rho
_{2}\in(0,1]$,
let $p$ be a positive integer, we have, as $n\rightarrow+\infty$,
\begin{eqnarray*} 
\mathop{\mathop{\mathop{\sup_{f_{i}=F(d_{1},g_{i})
f_{i}'=F(d_{2},g_{i}')}}
_{4p(d_{1}-d_{2})\leq\rho_{2}+1-t}}_{g_{i}\in\mathcal
{G}(-M,M,L,\rho_{1})}}
_{g_{i}'\in\mathcal{G}(m,M,L,\rho_{2})}
\Biggl|\frac{1}{n}\tr\Biggl[\prod
_{i=1}^{p}T_{n}(f_{i})T_{n}(f_{i}')^{-1}\Biggr]
-\frac{1}{2\pi}\int_{-\pi}^{\pi}\prod_{i=1}^{p}\frac
{f_{i}(\lambda
)}{f_{i}'(\lambda)}\,d\lambda\Biggr|\rightarrow0.
\end{eqnarray*}
%
\end{lemma}
\begin{pf} This result is a direct consequence
of Lemma~\ref{lemunifTR1}, as in \citet{jud03}. The only
difference is
in the proof of Lemma~5.2. of \citet{dhl89}, that is, in the study of
terms in the form
\[
|\id-T_{n}(f)^{1/2}T_{n}((4\pi^{2}f)^{-1})T_{n}(f)^{1/2}|
\]
with $f=F(d_{2},g_{i}')$ for any $i\leq p$. For simplicity's sake
we write $f=F(d,g)$ in the following calculations. Following Dahlhaus's
(\citeyear{dhl89}) proof, we obtain an upper bound of
$
|f(\lambda_{1})/f(\lambda_{2})-1|
$
which is different from \citet{dhl89}. If $g\in{\cal G}(m,M,L,\rho_{2})$,
the Lipschitz condition in $\rho_{2}$ implies that
\[
\biggl|\frac{f(\lambda_{1})}{f(\lambda_{2})}-1\biggr|\leq K
\biggl(|\lambda
_{1}-\lambda_{2}|^{\rho_{2}}+\frac{|\lambda_{1}-\lambda
_{2}|^{1-\delta
}}{|\lambda_{1}|^{1-\delta}}\biggr).
\]
Calculations as in Lemma 5.2 of \citet{dhl89} imply that
\[
|I-T_{n}(f)^{1/2}T_{n}((4\pi^{2}f)^{-1}
)T_{n}(f)^{1/2}|^{2}=O(n^{1-\rho_{2}}\log{n}^{2})+O(n^{\delta
})\qquad
\forall\delta>0.
\]
From this we prove the lemma following Lieberman, Rosemarin and
Rousseau [(\citeyear{LRR09}), Lemma 7], the
bounds being uniform over the considered class of functions.
\end{pf}

\subsection{\texorpdfstring{Approximations: Lemmas \protect\ref{lemLPunif} and
\protect\ref{whittles}}
{Approximations: Lemmas 5 and 6}}
\label{secALP}

We now propose a generalization of \citet{LP04}, whose proof is
given in
the supplementary material; see Lemma 1, Section 3, in \citet{RouChoLis}.
%
%
\begin{lemma} \label{lemLPunif} Let $1/2>a>0$, $L>0$, $M>0$ and $0<
\rho\leq1$.
Then for all $\delta>0$, there exists $C>0$
such that for all $n\in\N^{*}$,
%
%
\begin{eqnarray}
&&
\mathop{\sup_{p(d_{1}+d_{2})\leq a}}_{g_{j},g_{j}' \in\mathcal
{G}(-M,M,L,\rho)}
\Biggl|\frac{1}{n}\tr\Biggl[\prod
_{j=1}^{p}T_{n}(F(d_{1},g_{j}))T_{n}(F(d_{2},g_{j}'))\Biggr]
\nonumber\\
&&\hspace*{59pt}\qquad{} -(2\pi)^{2p-1}\int_{-\pi}^{\pi}\prod
_{j=1}^{p}F(d_{1},g_{j})F(d_{2},g_{j}')(x)\,dx \Biggr|
\\
&&\qquad
\leq Cn^{-\rho+\delta+2a_+},\nonumber
\end{eqnarray}
where $d_{1},d_{2}>-1/2$ and $a_{+}=\max(a,0)$.
\end{lemma}
%
%
\begin{lemma}\label{whittles} Let $f_{j}$, $j\in\{1,2\}$ be such
that $f_{j}(\lambda)=F(d_{j},g_{j})$, where $d_{j}\in(-1/2,1/2)$,
$0<m\leq g_{j}\leq M<+\infty$ for some positive constant $m,M$ and
consider $b$ a bounded function on $[-\pi,\pi]$. Assume that
$|d_{1}-d_{2}|<\delta$,
with $\delta\in(0,1/4)$; then, provided $d_{1}>d_{2}$, $\forall a >2
\delta$,
%
%
\begin{eqnarray}\label{whit0}
&&\frac{1}{n}\tr
[T_{n}(f_{1})^{-1}T_{n}(f_{1}b)T_{n}(f_{2})^{-1}T_{n}(f_{1}b)
]\nonumber\\[-8pt]\\[-8pt]
&&\qquad\leq
C(\log n)[|b|_{2}^{2}+(\delta+ n^{-1+6a}) |b|_{\infty}^{2}]\nonumber
\end{eqnarray}
and, without assuming $d_{1}>d_{2}$,
%
%
\begin{eqnarray}\label{eqwhit2}
&&\frac{1}{n}\tr
[T_{n}(f_{1}^{-1})T_{n}(f_{1}-f_{2})T_{n}(f_{2}^{-1})T_{n}(f_{1}-f_{2})]
\nonumber\\[-8pt]\\[-8pt]
&&\qquad \leq
C\bigl[h_{n}(f_{1},f_{2})+n^{\delta-1/2}\sqrt
{h_{n}(f_{1},f_{2})}\bigr].
\nonumber
\end{eqnarray}
\end{lemma}

\section{\texorpdfstring{Construction of tests: Lemmas \protect\ref{7}, \protect\ref{8} and \protect\ref{9}}
{Construction of tests: Lemmas 7, 8 and 9}}
\label{apptests}

%
%
\begin{lemma} \label{7} If $8|d_{0}-d_{i}|\leq\rho+1-t$
[case \textup{(a)} of condition (1)], the inequalities in (\ref{eqEphi}) are
verified provided $\rho_{i}=\tr[\id
-T_{n}(f_{0})T_{n}^{-1}(f_{i})]/n+h_{n}(f_{0},f_{i})$,
$f\leq f_{i}$ and
%
%
\begin{equation}\label{eqcondlema}
\frac{1}{2\pi}\int_{0}^{\pi}\frac{f_{i}(\lambda)-f(\lambda
)}{f_{0}(\lambda)}\,
d\lambda\leq h(f_{0},f_{i})/4.
\end{equation}
\end{lemma}
\begin{pf}
For all $s\in(0,1/4)$, using Markov inequality,
\begin{eqnarray*}
E_{0}^{n}[\phi_{i}] & \leq& \exp\{ -sn\rho_{i}\}
E_{0}^{n}\bigl[\exp\bigl\{ -s\Xn^{t}\{
T_{n}^{-1}(f_{i})-T_{n}^{-1}(f_{0})\} \mathbf{X}_{n}\bigr\}
\bigr]\\
& = & \exp\bigl\{ -sn\rho_{i}-\tfrac{1}{2}\log\det[\id
+2sB(f_{0},f_{i})]\bigr\} \\
& \leq& \exp\bigl\{ -sn\rho_{i}-s\tr[B(f_{0},f_{i})]\\
&&\hspace*{19.5pt}{}+s^{2}\tr\bigl[\bigl(\bigl(\id+2s\tau
B(f_{0},f_{i})\bigr)^{-2}B(f_{0},f_{i})\bigr)^{2}\bigr]\bigr\} \\
& \leq& \exp\{ -sn\rho_{i}-s\tr[B(f_{0},f_{i})
]+4s^{2}\tr[B(f_{0},f_{i})^{2}]\} ,
\end{eqnarray*}
where $\tau\in(0,1)$, using a Taylor expansion of the log-determinant
around $s=0$, and the following inequality:
\begin{eqnarray*}
&&
\id+2s\tau B(f_{0},f_{i})\\
&&\qquad=(1-2s\tau)\id+2s\tau
T_{n}(f_{0})^{1/2}T_{n}(f)^{-1}T_{n}(f_{0})\\
&&\qquad\geq\tfrac{1}{2}\id,
\end{eqnarray*}
since $s\tau<1/4$. Substituting $\rho_{i}$ with its expression,
the polynomial above is minimal for $s_{\min
}=h_{n}(f_{0},f_{i})/8b_{n}(f_{0},f_{i})$.
According to $s_{\min}\in(0,1/4)$ or not, that is, whether
$h_{n}(f_{0},f_{i})<2b_{n}(f_{0},f_{i})$
or not, one has
%
%
\begin{eqnarray}\label{eqE0phi}
\frac{1}{n} \log E_{0}^{n}[\phi_{i}]
& \leq& -\frac{h_{n}(f_{0},f_{i})^{2}}{16b_{n}(f_{0},f_{i})}\I
\{
h_{n}(f_{0},f_{i})<2b_{n}(f_{0},f_{i})\} \nonumber\\
& &{} -\frac{h_{n}(f_{0},f_{i})-b_{n}(f_{0},f_{i})}{4}\I\{
h_{n}(f_{0},f_{i})\geq2b_{n}(f_{0},f_{i})\} \\
& \leq& -\frac{h_{n}(f_{0},f_{i})}{16}\min\biggl\{ \frac
{h_{n}(f_{0},f_{i})}{b_{n}(f_{0},f_{i})},2\biggr\} .\nonumber
\end{eqnarray}
Since $8|d_{0}-d_{i}|\leq\rho+1-t$, the convergences
$b_{n}(f_{0},f_{i})\rightarrow b(f_{0},f_{i})$
and $h_{n}(f_{0},\allowbreak f_{i})\rightarrow h(f_{0},f_{i})$ are uniform on
the support of the prior $\pi$; see Lemma~\ref{lemhnbn}.
One deduces that, for any $a>0$
and $n$ large enough,
\[
\frac{1}{n} \log E_{0}^{n}[\phi_{i}]
\leq-\frac{n}{16}\min\biggl\{
\frac{h(f_{0},f_{i})^{2}-a}{b(f_{0},f_{i})+a},2h(f_{0},f_{i})-a
\biggr\}.
\]
Since $f_{i}\in\A_{\varepsilon}^{c}$, $h(f_0,f_i)>\varepsilon$, and
one may take $a=\varepsilon^{2}/2$ to obtain
\[
\frac{1}{n}\log E_{0}^{n}[\phi_{i}]
\leq-\frac{nh(f_{0},f_{i})}{32}\min\biggl\{ \frac
{h(f_{0},f_{i})}{b(f_{0},f_{i})+\varepsilon^{2}/2},2\biggr\}.
\]
Since $| d_0-d_i | \leq( \rho+1-t )/8\leq1/4
$, Lemma~\ref{ineqbh} (see Appendix~\ref{sectechlem}) implies
that
there exists $C_1>0$ such that
$E_{0}^{n}[\phi_{i}]\leq\exp{(-nC_{1}\varepsilon)}$
for $\varepsilon$ small enough.

If $f$ is in the support of $\pi$ and satisfies $f\leq f_{i}$,
and $8(d_{i}-d)\leq\rho+1-t$, using the same kind of calculations
and the fact that
\[
\id-2sT_{n}^{1/2}(f) \{ T_{n}^{-1}(f_{i})-T_{n}^{-1}(f_{0})
\} T_{n}^{1/2}(f)\geq\id+2sB(f,f_{0})
\]
as $T_{n}(f)\leq T_{n}(f_{i})$, we obtain for $s\in(0,1/4)$,
\begin{eqnarray*}
E_{f}^{n}[1-\phi_{i}] & \leq& \exp\{ ns\rho
_{i}-s\tr
[B(f,f_{0})]+4s^{2}\tr[B(f,f_{0})^{2}]
\} \\
& \leq& \exp\{ -nsh_{n}(f_{0},f_{i})+s\tr
[A(f_{i}-f,f_{0})]\\
&&\hspace*{91.6pt}{}+4s^{2}\tr[B(f,f_{0})^{2}]
\} \\
& \leq& \exp\{ -nsh_{n}(f_{0},f_{i})/2+4s^{2}\tr
[B(f,f_{0})^{2}]\},
\end{eqnarray*}
where the last inequality comes from (\ref{eqcondlema}), which
implies $\tr[A(f_{i}-f,f_{0})]/n\leq h_{n}(f_{0},f_{i})/2$
for $n$ large enough, uniformly in $f$, using Lemma~\ref{lemhnbn}. Doing the
same calculations as above, for $n$ large enough,
%
%
\begin{eqnarray} \label{secondtypecasea}
\frac{1}{n}\log E_{f}^{n}[1-\phi_{i}] & \leq& -\frac
{1}{64}\min\biggl\{ \frac
{h_{n}(f_{0},f_{i})^{2}}{b_{n}(f,f_{0})},4h_{n}(f_{0},f_{i})\biggr\}
\nonumber\\[-8pt]\\[-8pt]
& \leq& -\frac{1}{64}\min\biggl\{ \frac
{h(f_{0},f_{i})^{2}/2}{b(f,f_{0})+\varepsilon
^{2}/2},2h(f_{0},f_{i})\biggr\} .\nonumber
\end{eqnarray}
To conclude, note that $f\leq f_{i}$, and (\ref{eqcondlema}) implies
that
\begin{eqnarray*}
b(f,f_{0}) & = & \frac{1}{2\pi}\int_{-\pi}^{\pi}\biggl\{ \frac
{f^{2}}{f_{0}^{2}}+1-2\frac{f}{f_{0}}\biggr\} \, d\lambda\\
& \leq& b(f_{i},f_{0})+h(f_{0},f_{i})/2
\leq(C+1/2)h(f_{0},f_{i}),
\end{eqnarray*}
according to Lemma~\ref{ineqbh}. One concludes that there exists
$C_{1}>0$ such that $E_{f}^{n}[1-\phi_{i}]\leq
e^{-nC_{1}\varepsilon}$.
\end{pf}
%
%
\begin{lemma} \label{8} If $8(d_{i}-d_{0})>\rho+1-t$ [case
\textup{(b)} of condition (3)], the inequalities (\ref{eqEphi}) are verified
provided $\rho_{i}=\tr[\id-T_{n}(f_{0})T_{n}^{-1}(f_{i})
]/n+2\KL_{n}(f_{0};f_{i})$,
for any $f$ such that $f\leq f_{i}$ and
%
%
\begin{equation}\label{eqcondlemb}\qquad
\frac{1}{2\pi}\int_{-\pi}^{\pi} \biggl( \frac{f_{i}}{f}-1
\biggr)\,
d\lambda\leq\biggl(\frac{M}{\pi^{2}m}\biggr)^{4}\frac
{b(f_{0},f_{i})}{64},\qquad
b(f_{i},f)\leq b(f_{0},f_{i}).
\end{equation}
\end{lemma}

For $\varepsilon$ small enough, if $b(f_{i},f)\leq b(f_{0},f_{i})|{\log
\varepsilon}|^{-1}$,
(\ref{eqcondlemb}) is satisfied.
\begin{pf}
The upper bound of $E_{0}^{n}[\phi_{i}]$ is computed similarly to
(\ref{eqE0phi}) so that
\[
\frac{1}{n}\log E_{0}^{n}[\phi_{i}]\leq-\frac{1}{4}\min\biggl\{
\frac
{\KL_{n}(f_{0},f_{i})^{2}}{b_{n}(f_{0},f_{i})},\KL_{n}(f_{0},f_{i})
\biggr\}
.
\]
According to Lemma~\ref{lemdgebge} and since
$8(d_{i}-d_{0})\geq\rho+1-t$, there exists \mbox{$C>0$}, such that
$b(f_{0},f_{i})\geq C$. Using the uniform convergence results of
Appendix~\ref{sectechn-lemm-contr}, this means that $b_n(f_0,f_i)\geq
C/2$,
for $n$ large enough, independently of $f_i$. Using Lemma
\ref{lemdahlhaus}, there exists a constant $C_1\leq1$ such that
$\KL_{n}(f_{0},f_{i})\geq C_1b_{n}(f_{0},f_{i})$. Thus, there exists
$C_{2}>0$ such that
\[
\frac{1}{n} \log E_{0}^{n}[\phi_{i}]
\leq-nC_2b(f_{0},f_{i})
\]
and, for $\varepsilon$ small enough, and some $C_{3}>0$,
$
E_{0}^{n}[\phi_{i}]\leq\exp\{-nC_{3}\varepsilon\}$.

As in the previous lemma, let $h\in(0,1)$,
\begin{eqnarray*}
&&\log E_{f}^{n}[1-\phi_{i}]\\
&&\qquad \leq (1-h)n\rho_{i}/2\\
&&\qquad\quad{} -\tfrac{1}{2}\log\det[\id-(1-h)T_{n}(f)^{1/2}\{
T_{n}^{-1}(f_{i})-T_{n}^{-1}(f_{0})\} T_{n}(f)^{1/2}]\\
&&\qquad \leq (1-h)n\rho_{i}/2-\tfrac{1}{2}\log\det[\id
+(1-h)B(f,f_{0})]\\
&&\qquad = (1-h)n\rho_{i}/2-\log\det[A(f,f_{0})]/2\\
&&\qquad\quad{} -\tfrac{1}{2}\log\det[\id
(1-h)+hT_{n}^{-1/2}(f)T_{n}(f_{0})T_{n}^{-1/2}(f)].
\end{eqnarray*}
Substituting $\rho_{i}$ with its expression, that is,
$n\rho_{i}-\log\det A(f,f_{0})
= \log\det A(f_{i},f)$
and using the same kind of expansions as in the previous lemma, one
obtains
\begin{eqnarray*}
&&\frac{1}{n} \log E_{f}^{n}[1-\phi_{i}]\\
&&\qquad\leq \frac{1}{n} \log\det[A(f_{i},f)]+ (h/2)\tr[ T_n(f_0)
\{ T_n^{-1}(f_i) - T_n^{-1}(f) \} ] \\
&&\qquad\quad{} -hn \mathit{KL}_n(f_0;f_i) +h^2 \tr[ \{
\id-T_n^{-1}(f)T_n(f_{0}) \}^{2} ] \\
&&\qquad\leq \frac{1}{n} \log\det[A(f_{i},f)]  -hn \mathit{KL}_n(f_0;f_i)
+h^2 \tr[ \{
\id-T_n^{-1}(f)T_n(f_{0}) \}^{2} ] \\
&&\qquad\leq + \frac{1}{n} \log\det[A(f_{i},f)]
-n\min\biggl(\frac{\KL_{n}(f_{0},f_{i})^{2}}{4\tr
{B(f_{0},f)^{2}}/n},\frac{\KL_{n}(f_{0},f_{i})}{4}\biggr).
\end{eqnarray*}
Note that we use the fact $f\leq f_i$ in the second line.

Since $\log\det A(f_{i},f)=\log\det\{ \id
+T_{n}(f_{i}-f)T_{n}(f)^{-1}\} $,
using a Taylor expansion of $\log{\det}$ around $\id$, we obtain
that for $n$ large enough,
\[
\frac{1}{n}\log\det A(f_{i},f)\leq\frac{1}{2\pi}\int_{-\pi}^{\pi
}\frac
{f_{i}-f}{f}\,d\lambda+a,
\]
where $a$ can be chosen as small as necessary. In addition, we use
Lemma~\ref{lemdahlhaus} and the uniform convergence results of
Lemmas~\ref{lemunifTR1},~\ref{lemunifTR2} to obtain that
\[
\frac{(n\KL_{n}(f_{0},f_{i}))^{2}}{\tr[B(f_{0},f)^{2}]}\geq\frac
{nm^{4}(b(f_{0},f_{i})^{2}-a)^{2}}{16\pi^{8}M^{4}(b(f_{0},f)+a)}
\]
and, since $d\geq d_{0}$ and (\ref{eqcondlemb}),
\begin{eqnarray*}
b(f_{0},f) & = & \frac{1}{2\pi}\int_{-\pi}^{\pi}\biggl(\frac
{f_{0}}{f}-1\biggr)^{2}\, d\lambda
\leq 2\biggl(b(f_{0},f_{i})+\frac{M^{2}\pi
^{4}}{m^{2}}b(f_{i},f)\biggr)\\
&\leq& 2b(f_{0},f_{i})\biggl(1+\frac{M^{2}\pi^{4}}{m^{2}}\biggr);
\end{eqnarray*}
hence, under the constraint (\ref{eqcondlemb}), there exists $C_{1}>0$
such that, for $n$ large enough, $\varepsilon$ small enough,
$
E_{f}^{n}[1-\phi_{i}] \leq
\exp\{-nC_{1}b(f_{0},f_{i})\}\leq e^{-n\varepsilon}
$.
\end{pf}
%
%
\begin{lemma} \label{9} If $8(d_{0}-d_{i})>\rho+1-t$ [case
\textup{(c)} of condition (3)], the inequalities (\ref{eqEphi}) are verified
provided $\rho_{i}=\log{\det[T_{n}(f_{i})T_{n}(f_{0})^{-1}]}/n$ if
%
%
\begin{equation}\label{constlemcasec}\qquad
\frac{1}{2\pi}\int_{-\pi}^{\pi}\frac{f_{i}-f}{f_{0}}(\lambda
)\,d\lambda
\leq\frac{m^{2}}{4M^{2}\pi^{4}}b(f_{i},f_{0}),\qquad b(f,f_{i})\leq
b(f_{i},f_{0}).
\end{equation}
\end{lemma}

For $\varepsilon>0$ small if $\int(f_{i}-f)f_{i}^{-1}\,d\lambda\leq
b(f_{i},f_{0})|{\log\varepsilon}|^{-1}$,
(\ref{constlemcasec}) is satisfied.
\begin{pf}
For $0<h<1$, following the same lines as above, one has
\begin{eqnarray*}
\frac{1}{n}\log E_{0}^{n}[\phi_{i}] & \leq& -(1-h)n\rho
_{i}/2+\log{\det[A(f_{0},f_{i})]}/2\\
& &{} -\frac{1}{2}\log{\det[ \id
(1-h)+hT_{n}^{-1/2}(f_{0})T_{n}(f_{i})T_{n}^{-1/2}(f_{0}) ]} \\
& \leq&
-nh\KL_{n}(f_{i},f_{0})+h^{2}\tr{[B(f_{i},f_{0})^{2}]}
\leq-\varepsilon.
\end{eqnarray*}
Moreover, for all $f\leq f_{i}$, satisfying $8(d_{i}-d)\leq\rho+1-t$,
using the same calculations as in the proof of Lemma~\ref{7},
we bound $\log E_{f}^{n}[1-\phi_{i}]$ by the maximum
of the two following quantities:
\begin{eqnarray*}
&\displaystyle -\frac{ \{ n\KL_{n}(f_{i},f_{0})-\tr[A(f_{i}-f,f_{0})]/2
\}
^2}{4n\{b(f,f_{0})+a\}},&\\
&\displaystyle -\frac{n}{4}\KL_{n}(f_{i},f_{0})+ \frac{1}{8}
\tr[A(f_{i}-f,f_{0})],&
\end{eqnarray*}
where $a$ is any positive constant and $n$ is large enough. Using
Lemma~\ref{lemdahlhaus}, one has
\[
n\KL_{n}(f_{i},f_{0})\geq\frac{nm^{2}}{2\pi^{4}M^{2}}b(f_{i},f_{0}),
\]
and the constraints (\ref{constlemcasec}) we finally obtain that
there exists constant \mbox{$c_{1},C_{1}>0$} such that, for $\varepsilon$
small enough,
\begin{eqnarray*}
E_{f}^{n}[1-\phi_{i}] & \leq& \exp\bigl\{
-2n\bigl(\KL_{n}(f_{i},f_{0})-\tr[A(f_{i}-f,f)]/2n\bigr)+4s^{2}nb_{n}(f,f_{0})\bigr\}\\
& \leq& e^{-nc_{1}b(f_{i},f_{0})}\leq e^{-nC_{1}\varepsilon}.
\end{eqnarray*}
\upqed\end{pf}

\section{\texorpdfstring{Proof of Theorem \protect\ref{42}}{Proof of Theorem 4.2}}
\label{secproof-theor-rates}

We re-use some of the notation of Section
\ref{appprthcons}; in particular, $C$, $C'$ denote generic
constants.

The proof of the theorem is divided in two parts. First, we show that
%
%
\begin{equation}\label{fexprate1}
E_{0}^{n}\biggl[P^{\pi}\biggl\{ f\dvtx h_{n}(f,f_{0})\geq
\frac{\log
n}{n^{2\beta/(2\beta+1)}}\Big|\mathbf{X}_{n}\biggr\}
\biggr]
\leq\frac{C}{n^{2}}.
\end{equation}
Second, we show that, for $f\in\bar{\F}_n$, and $n$ large enough,
%
%
\begin{equation}\label{fexprate2}
h_{n}(f,f_{0})\leq
Cn^{-{2\beta}/({2\beta+1})}\log{n}\quad\Rightarrow\quad
h(f,f_{0})\leq
C'n^{-{2\beta}/({2\beta+1})} \log n.\hspace*{-35pt}
\end{equation}

Since $\ell(f,f_{0})\leq h(f,f_{0})$ (see
the proof of Corollary 2 in the supplementary material [\citet{RouChoLis}]),
the right-hand side inequality of (\ref{fexprate2}) implies that
\begin{eqnarray*}
E_{0}^{n}\{ E^{\pi}[\ell(f,f_{0})|\Xn]\} &
\leq&
C\frac{\log n}{n^{2\beta/(2\beta+1)}}\\
& &{} +\bar{\ell}E_{0}^{n}\biggl\{ P^{\pi}\biggl(h_{n}(f,f_{0})>\frac
{\log
n}{n^{2\beta/(2\beta+1)}}\Big|\Xn\biggr)\biggr\} \\
& \leq& Cn^{-{2\beta}/({2\beta+1})}\log n+C'n^{-2}
\end{eqnarray*}
for large $n$, where $\bar{\ell}<+\infty$ is an upper bound for
$\ell(f,f_{0})$ which is easily deduced from the fact that $f$,
$f_{0}$ belongs to some Sobolev class of functions. This implies
Theorem~\ref{42}.

To prove (\ref{fexprate1}), we show that conditions~\ref{ratecond1}
and~\ref{ratecond2} of Theorem~\ref{thratespde} are fulfilled
for
$
u_{n}=n^{-2\beta/(2\beta+1)}(\log n)$. In order to establish condition
\ref{ratecond1}, we show that, for
$n$ large enough, $\bar{\mathcal{B}}_{n}\supset\hat{\mathcal{B}}_{n}$,
the set containing all the $f=\tilde{F}(d,k,\theta)$ such that $k\geq
\bar{k}_{n}$,
for $\bar{k}_{n}=k_{0}n^{1/(2\beta+1)}$, $d-u_{n}n^{-a}\leq d_{0}\leq d$
and, for $j=0,\ldots,k$,
%
%
\begin{equation}\label{eqthetajminthetaj0}
|\theta_{j}-\theta_{0j}|\leq(j+1)^{-2\beta}u_{n}n^{-a},
\end{equation}
where\vspace*{1pt} $a>0$ is some small constant. Then it is easy to see that
$\pi(\bar{\mathcal{B}}_{n})\geq\pi(\hat{\mathcal{B}}_{n})\geq
\exp\{
-nu_{n}/2\}$,
provided $k_{0}$ is small enough, since $\pi_{k}(k\geq\bar
{k}_{n})\geq
\exp\{-C\bar{k}_{n}\log\bar{k}_{n}\}$,
and (\ref{eqthetajminthetaj0}) for all $j$ implies that
\begin{eqnarray*}
\sum_{j=0}^{k}\theta_{j}^{2}(j+1)^{2\beta} & = & \sum
_{j=0}^{k}(\theta
_{0j}-\theta_{0j}+\theta_{j})^{2}(j+1)^{2\beta}\\
& \leq& L_{0}+u_{n}^{2}n^{-2a}\sum_{j=0}^{k}(1+j)^{-2\beta
}+2u_{n}n^{-a}\Biggl(\sum_{j=1}^{k}|\theta_{0j}|\Biggr)\\
& < & L
\end{eqnarray*}
for $n$ large enough, since $L_{0}=\sum_{j}\theta_{0j}(j+1)^{2\beta}<L$,
and $\sum_{j=1}^{k}|\theta_{0j}|$ is bounded according to (\ref
{eqgeneralinequalitysum}).

Let $f=\tilde{F}(d,k,\theta)$, with $(d,k,\theta)\in\hat{\mathcal
{B}}_{n}$.
To prove that $(d,k,\theta)\in\bar{\mathcal{B}}_{n}$, it is sufficient
to prove that $h_{n}(f,f_{0})\leq u_{n}/4$, since
$h_{n}(f,f_{0})=\KL_{n}(f_{0};f)+\KL_{n}(f;f_{0})$,
and $\KL_{n}(f;f_{0})\geq Cb_{n}(f_{0},f)$, using the same calculation
as in Dahlhaus [(\citeyear{dhl89}), page 1755] and the fact that $d\leq d_{0}$.

Since $f_{0}\in\So(\beta,L)$, and for the particular choice of $\bar{k}_{n}$
above,
%
%
\begin{equation}\label{eqtheta0l2}
\sum_{j=\bar{k}_{n}}^{+\infty}\theta_{0j}^{2}\leq L(\bar
{k}_{n}+1)^{-2\beta}
\end{equation}
and
\[
\sum_{j=\bar{k}_{n}}^{+\infty}|\theta_{0j}|
\leq\Biggl(\sum_{j=\bar{k}_{n}}^{+\infty}\theta
_{0j}^{2}(j+1)^{2\beta
}\Biggr)^{1/2}\Biggl(\sum_{j=\bar{k}_{n}}^{+\infty}(j+1)^{-2\beta
}\Biggr)^{1/2}
\leq C\bar{k}_{n}^{1/2-\beta}.
\]
Let
\begin{eqnarray}\label{eqtheta0l}
f_{0n}(\lambda) & = & |1-e^{i\lambda}|^{-2d_{0}}\exp\Biggl(\sum
_{j=0}^{\bar{k}_{n}}\theta_{0j}\cos(j\lambda)\Biggr),\nonumber\\[-8pt]\\[-8pt]
b_{n}(\lambda) & = & \exp\biggl(-\sum_{j\geq\bar{k}_{n}+1}\theta
_{0j}\cos
(j\lambda)\biggr)-1\nonumber
\end{eqnarray}
and $g_{n}=1-f_{0n}/f$. Then $f-f_{0}=f_{0}b_{n}+fg_{n}$, where
$b_{n}$ and $g_{n}$ are bounded as follows. From (\ref{eqtheta0l}),
one gets that, for $n$ large enough, $|b_{n}|_{\infty}\leq C\bar
{k}_{n}^{1/2-\beta}$,
and
\[
|b_{n}|_{2}^{2}=\int_{-\pi}^{\pi}b_{n}(\lambda)^{2}\, d\lambda\leq
2\sum
_{j=\bar{k}_{n}+1}^{\infty}\theta_{0j}^{2}\leq2L\bar
{k}_{n}^{-2\beta
}\leq2Lk_{0}^{-2\beta}\frac{u_{n}}{\log n}
\]
according to (\ref{eqtheta0l2}). In addition since $1-x\leq-\log x$,
for $x>0$,
\begin{eqnarray*}
g_{n}(\lambda) & \leq& (d_{0}-d)\log(1-\cos\lambda)+\sum_{j\leq
\bar
{k}_{n}}|\theta_{0j}-\theta_{j}|\\
& \leq& Cu_{n}n^{-a}(|{\log}|\lambda||+1).
\end{eqnarray*}
Moreover, since $\tr\{ (A+B)^{2}\} \leq2\tr A^{2}+2\tr B^{2}$
for square matrices $A$ and~$B$, one has
\begin{eqnarray*}
h_{n}(f_{0},f) & \leq& \frac{1}{n}\tr
[T_{n}(f_{0}b_{n})T_{n}^{-1}(f)T_{n}(f_{0}b_{n})T_{n}^{-1}(f_{0})
] \\
& &{} +\frac{1}{n}\tr
[T_{n}(fg_{n})T_{n}^{-1}(f)T_{n}(fg_{n})T_{n}^{-1}(f_{0})]\\
& \leq& C\log n\{ |b_{n}|_{2}^{2}+u_{n}n^{-a}|b_{n}|_{\infty
}^{2}\} \\
& &{} +Cu_{n}^{2}n^{-1-2a}\tr\bigl[\bigl(T_{n}\bigl(f(|{\log}|\lambda
||+1)\bigr)T_{n}^{-1}(f)\bigr)^{2}\bigr]\\
&\leq& cu_{n},
\end{eqnarray*}
where $c$ may be chosen as small as necessary, since $k_{0}$ is
arbitrarily large. Note that the first two terms above come from
(\ref{whit0}) in Lemma~\ref{whittles}, and the third term comes
from Lemma~\ref{lemunifTR2}.

To establish condition~\ref{ratecond2} is straightforward, since
the prior has the same form as in Section~\ref{secconsfexp}, and
we can use the same reasoning as in the proof of Theorem~\ref{thmfexp-conv};
that is, we take, for some suitably chosen $\delta$,
\[
\bar{{\cal F}}_{n}=\{ (d,k,\theta)\in\So(\beta,L)\dvtx
|d-d_{0}|\leq
\delta, k\leq\tilde{k}_{n}\},
\]
where $\tilde{k}_{n}=k_{1}n^{1/(2\beta+1)}$
so that, using Lemma~\ref{lemdgehge},
\[
\pi\bigl(\bar{\mathcal{F}}_{n}^{c}\cap\{f,h(f,f_{0})<\varepsilon\}
\bigr)\leq\pi_{k}(k\geq\tilde{k}_{n})\leq e^{-C\tilde{k}_{n}\log
{\tilde{k}_{n}}}
\]
for $n$ large enough. Choosing $k_{1}$ large enough leads to
condition\vadjust{\goodbreak}
(2).

We now verify condition~\ref{ratecond3} of Theorem~\ref{42}.
Let $\varepsilon_{n}^{2}\geq u_{n}$ and $l_{0}\leq l\leq l_{n}$,
and consider $f=\tilde{F}(d,k,\theta)$, $(d,k,\theta)\in
\mathcal{V}_{n,l}$,
as defined in Theorem~\ref{thratespde}, and
$f_{i,l}=(2e)^{l\varepsilon_{n}^{2}}\tilde{F}(d_{i},k,\theta_{i})$,
where dependencies on $l$ in $d_{i}$ and $\theta_{i}$ are dropped
for convenience. If for some positive $c>0$ to be chosen accordingly
$|\theta_{j}-\theta_{ij}|\leq cl\varepsilon_{n}^{2}/(k+1)$, for
$j=0,\ldots,k$,
one obtains
\[
\frac{g_{i,l}(\lambda)}{g(\lambda)} = (2e)^{l\varepsilon
_{n}^{2}}\exp
\Biggl\{ \sum_{j=0}^{k}(\theta_{j}-\theta_{ij})\cos(j\lambda
)\Biggr\}
\leq(2e^{2})^{cl\varepsilon_{n}^{2}}
\]
and $f_{i,l}/f\geq1$ so that the constraints of condition \ref
{ratecond3} of Theorem~\ref{42}
are verified by choosing $c$ small enough.
The cardinal of the smallest possible net under these constraints
needed to cover $\mathcal{V}_{n,l}$ is bounded by
\[
\bar{\mathcal{C}}_{n,l}\leq k_{n}\biggl(\frac{1}{cl\varepsilon
_{n}^{2}}\biggr)\biggl(\frac{L'k_{n}}{cl\varepsilon_{n}^{2}}
\biggr)^{k_{n}+1}
\]
since for all $l$ $|\theta_{l}|\leq L$. This implies that
$\log\bar{\mathcal{C}}_{n,l}\leq Cnu_{n}$,
and condition~\ref{ratecond3} is verified with $\varepsilon
_{n}^{2}=\varepsilon_{0}^{2}u_{n}$.
This achieves the proof of (\ref{fexprate1}), which provides a~rate of convergence in terms of the distance $h_{n}(\cdot,\cdot)$.

Finally, we prove (\ref{fexprate2}) to obtain a rate of convergence
in terms of the distance $h(\cdot,\cdot)$. Consider $f$ such that
\[
h_{n}(f_{0},f) = \frac{1}{2n}\tr
[T_{n}^{-1}(f_{0})T_{n}(f-f_{0})T_{n}^{-1}(f)T_{n}(f-f_{0})]\leq
\varepsilon_{n}^{2}.
\]
Equation (\ref{eqwhit2}) of Lemma~\ref{whittles} implies that
%
%
\begin{eqnarray}\label{eqexD7}
&&\frac{1}{2n}\tr
[T_{n}(f_{0}^{-1})T_{n}(f-f_{0})T_{n}(f^{-1})T_{n}(f-f_{0})]\nonumber\\[2pt]
&&\qquad\leq
C\varepsilon_{n}[\varepsilon_{n}+n^{-1/2+\delta}]\\[2pt]
&&\qquad \leq C\varepsilon_{n}^{2}.\nonumber
\end{eqnarray}
We now prove that
\begin{eqnarray*}
&&\tr
[T_{n}(f_{0}^{-1})T_{n}(f-f_{0})T_{n}(f^{-1})T_{n}(f-f_{0})]\\[2pt]
&&\quad{}-\tr\bigl[T_{n}\bigl(f_{0}^{-1}(f-f_{0})\bigr)T_{n}\bigl(f^{-1}(f-f_{0})\bigr)\bigr] \\[2pt]
&&\qquad
\leq \frac{C(\log n)^{2}}{n^{1-2a}}
\end{eqnarray*}
for some small $a>0$. By symmetry we consider only the case $d\geq d_{0}$.
Let $h_{0}=(1-\cos\lambda)^{d_{0}}$, $h=(1-\cos{\lambda})^{d}$,
then $fh\leq C$, $f_{0}h_{0}\leq C$ and $|f-f_{0}|h\leq C$ for
some $C\geq0$, and it is sufficient to study the difference below.
Note that the calculations below follow the same lines and the same
notation as the treatment\vadjust{\goodbreak} of $\gamma(b)$ in Lemma~\ref{whittles}; see
Appendix~\ref{sectechn-lemm-contr}; in particular, $\Delta_n(\lambda)
=\sum_{j=1}^n \exp(-i\lambda j)$, and $L_n(\lambda)=n$ for $|\lambda
|\leq1/n$,
$L_n(\lambda)=|\lambda|^{-1}$ otherwise.
\begin{eqnarray*}
&&\frac{1}{n}\tr
\bigl[T_{n}\bigl(h_{0}(f-f_{0})\bigr)T_{n}\bigl(h(f-f_{0})\bigr)\bigr]\\
&&\quad{}-\frac{1}{n}\tr
[T_{n}(h_{0})T_{n}(f-f_{0})T_{n}(h)T_{n}(f-f_{0})]\\
&&\qquad = -\frac{1}{n}\int_{[-\pi,\pi]^{3}}(f-f_{0})(\lambda
_{2})h_{0}(\lambda_{2})(f-f_{0})(\lambda_{4})h(\lambda_{4})
\biggl(\frac
{h_{0}(\lambda_{1})}{h_{0}(\lambda_{2})}-1\biggr)\\
&&\hspace*{52pt}\qquad\quad{} \times\Delta_{n}(\lambda_{1}-\lambda_{2})\Delta
_{n}(\lambda
_{2}-\lambda_{4})\Delta_{n}(\lambda_{4}-\lambda_{1})\,d\bla\\
&&\qquad\quad{} -\frac{1}{n}\int_{[-\pi,\pi]^{3}}(f-f_{0})(\lambda
_{2})h_{0}(\lambda_{1})(f-f_{0})(\lambda_{4})h(\lambda_{4})
\biggl(\frac
{h(\lambda_{3})}{h(\lambda_{4})}-1\biggr)\\
&&\hspace*{56pt}\qquad\quad{} \times\Delta_{n}(\lambda_{1}-\lambda_{2})\Delta
_{n}(\lambda
_{2}-\lambda_{3})\Delta_{n}(\lambda_{3}-\lambda_{4})\Delta
_{n}(\lambda
_{4}-\lambda_{1})\,d\bla\\
&&\qquad \leq \frac{C(\log n)}{n}\int_{[-\pi,\pi]^{2}}|\lambda
_{2}|^{-2(d-d_{0})}|\lambda_{1}|^{-1+a}L_{n}(\lambda_{1}-\lambda
_{2})^{1+a}\,d\bla\\
&&\qquad\quad{} +\frac{C}{n}\int_{[-\pi,\pi]^{4}}\frac{|\lambda
_{1}|^{2d}}{|\lambda
_{2}|^{2d}|\lambda_{3}|^{1-a}}
L_{n}(\lambda_{1}-\lambda_{2})L_{n}(\lambda
_{2}-\lambda_{3})\\
&&\hspace*{60pt}\qquad\quad{} \times L_{n}(\lambda_{3}-\lambda_{4})^{a}L_{n}(\lambda_{4}-\lambda
_{1})\,d\bla\\
&&\qquad \leq \frac{C(\log n)^{2}}{n^{1-a}}\int_{[-\pi,\pi]^{2}}|\lambda
_{2}|^{-2(d-d_{0})}|\lambda_{1}|^{-1+a}L_{n}(\lambda_{2}-\lambda
_{1})\,d\bla\\
&&\qquad\quad{} +\frac{C(\log n)}{n^{1-a}}\int_{[-\pi,\pi]^{3}}\frac{|\lambda
_{1}|^{2d}}{|\lambda_{2}|^{2d}|\lambda_{3}|^{1-a}}L_{n}(\lambda
_{1}-\lambda_{2})L_{n}(\lambda_{2}-\lambda_{3})\,d\bla\\
&&\qquad \leq \frac{C(\log n)^{2}}{n^{1-2a}},
\end{eqnarray*}
provided $d-d_{0}\leq a/4$, using standard calculations.
Combined with (\ref{eqexD7}),
this result implies that
\[
\frac{1}{n}\tr\bigl[T_{n}\bigl(h_{0}(f-f_{0})\bigr)T_{n}
\bigl(h(f-f_{0})\bigr)\bigr]\leq C\epsilon_{n}^{2}.
\]
Finally, to obtain (\ref{fexprate2}), we bound
\begin{eqnarray*}
&&\bigl|\tr
\bigl[T_{n}\bigl(h_{0}(f-f_{0})\bigr)T_{n}\bigl(h(f-f_{0})\bigr)
\bigr]-\tr\bigl[T_{n}\bigl(h_{0}h(f-f_{0})^{2}\bigr)\bigr]\bigr|\\
&&\qquad = C\biggl|\int_{[-\pi,\pi]^{2}}\{ h_{0}(f-f_{0})\}
(\lambda_{1})[\{ h(f-f_{0})\} (\lambda_{2})-\{
h(f-f_{0})\} (\lambda_{1})]\\
&&\hspace*{150pt}\qquad\quad{} \times\Delta_{n}(\lambda
_{1}-\lambda
_{2})\Delta_{n}(\lambda_{2}-\lambda_{1})\,d\bla\biggr|\\
&&\qquad \leq C\biggl|\int_{[-\pi,\pi]^{2}}\{ h(f-f_{0})\}
(\lambda_{1})(f-f_{0})(\lambda_{2})[h(\lambda_{2})-h(\lambda
_{1})]\\
&&\hspace*{108.2pt}\qquad\quad{}\times\Delta
_{n}(\lambda_{1}-\lambda_{2})\Delta_{n}(\lambda_{2}-\lambda
_{1})\,d\bla
\biggr|\\
&&\qquad\quad{} +C\biggl|\int_{[-\pi,\pi]^{2}}\{ hh_{0}(f-f_{0})\}
(\lambda_{1})[f_{0}(\lambda_{2})-f_{0}(\lambda_{1})
]\\
&&\hspace*{82.5pt}\qquad\quad{}\times\Delta
_{n}(\lambda_{1}-\lambda_{2})\Delta_{n}(\lambda_{2}-\lambda
_{1})\,d\bla
\biggr|\\
&&\qquad\quad{} +C\biggl|\int_{[-\pi,\pi]^{2}}\{ hh_{0}(f-f_{0})\}
(\lambda_{1})[f(\lambda_{2})-f(\lambda_{1})]\\
&&\hspace*{77pt}\qquad\quad{}\times\Delta
_{n}(\lambda_{1}-\lambda_{2})\Delta_{n}(\lambda_{2}-\lambda
_{1})\,d\bla
\biggr|.
\end{eqnarray*}
The first term is of order $O(n^{2a}\log{n})$, from the same calculations
as above. We consider the last term, but the calculations for the
second term follow exactly the same lines. Recall that $f=he^{w}$,
where $w(\lambda)=\sum_{j=0}^{k}\theta_{j}\cos(j\lambda)$ is not
necessarily continuously differentiable, for example, when $\beta<1$. Thus
\begin{eqnarray*}
f(\lambda_{2})-f(\lambda_{1})&=&[h(\lambda_{2})^{-1}-h(\lambda
_{1})^{-1}]e^{w(\lambda_{2})}\\
&&{}+h(\lambda_{1})^{-1}
\bigl[e^{w(\lambda_{2})}-e^{w(\lambda_{1})}\bigr].
\end{eqnarray*}
The first term is dealt with using (5) and (6) in the supplementary
material [\citet{RouChoLis}],
leading to a
bound of order $(\log n)^{2}n^{2a}$. For the second term, and $k\leq
k_{n}$,
\begin{eqnarray*}
&&\biggl|\int_{[-\pi,\pi]^{2}}h_{0}(f-f_{0})(\lambda
_{1})[g(\lambda_{2})-g(\lambda_{1})]\Delta_{n}(\lambda_{1}-\lambda
_{2})\Delta_{n}(\lambda_{2}-\lambda_{1})\,d\bla\biggr|\\
&&\qquad \leq C\int_{[-\pi,\pi]^{2}}h_{0}|f-f_{0}|(\lambda
_{1})
\Biggl|\sum_{j=0}^{k}\theta_{j}\bigl(\cos(j\lambda_{2})-\cos(j\lambda
_{1})\bigr)\Biggr|\\
&&\qquad\quad\hspace*{44.5pt}{}\times L_{n}(\lambda_{1}-\lambda_{2})L_{n}(\lambda
_{2}-\lambda_{1})\,d\bla\\
&&\qquad \leq C(\log n)\Biggl(\sum_{j=0}^{k}|\theta_{j}|j\Biggr)\int
_{-\pi
}^{\pi}\{ h_{0}|f-f_{0}|\} (\lambda_{1})\,d\lambda_{1}\\
&&\qquad \leq C(\log n)\Biggl(\sum_{j=0}^{k}|\theta_{j}|j\Biggr)
\biggl(\int
_{-\pi}^{\pi}\{ hh_{0}(f-f_{0})^{2}\} (\lambda)\,d\bla
\biggr)^{1/2},
\end{eqnarray*}
where the latter inequality holds because $\int_{-\pi}^{\pi}\{
h_{0}/h\} (\lambda)\,d\bla$
is bounded when $|d-d_{0}|$ is small enough. The same computations
can be made on $f_{0}$ so that for all $a>4|d-d_{0}|$, we finally
obtain that
\begin{eqnarray*}
\hspace*{-5.5pt}&&\bigl|\tr
\bigl[T_{n}\bigl(h_{0}(f-f_{0})\bigr)T_{n}\bigl(h(f-f_{0})\bigr)
\bigr]-\tr\bigl[T_{n}\bigl(h_{0}h(f-f_{0})^{2}\bigr)\bigr]\bigr|\\
\hspace*{-5.5pt}&&\qquad \leq C(\log n)n^{2a}+(\log n)\sum_{j=0}^{k}j(|\theta_{j}|+|\theta
_{0j}|)\biggl(\int_{[-\pi,\pi]}g_{0}g(f-f_{0})^{2}(\lambda)\,d\lambda
\biggr)^{1/2}.
\end{eqnarray*}
Splitting the indices of the sum above into into $\{ j\dvtx j|\theta
_{j}|\leq j^{2\beta+r}\theta_{j}^{2}\} $
and its complementary, for some $r$, we get that
\[
\sum_{j=0}^{k}j|\theta_{j}| \leq\sum_{j=0}^{k}j^{2\beta+r}\theta
_{j}^{2}+\sum_{j=0}^{k}j^{1-2\beta-r}
\leq C(k^{r}+k^{2-2\beta-r})\leq Ck_{n},
\]
provided\vspace*{1pt} we take $r=3/2-\beta$. One concludes by doing the same
computation for~$f_{0}$,
so as to obtain that, for $\beta\geq1/2$,
$
\int_{-\pi}^\pi h_{0}h(f_{0}-f)^{2}\,d\lambda\leq C\varepsilon_{n}^{2}$.

\section{Technical lemmas}
\label{sectechlem}

The three following lemmas provide inequalities involving
\[
b(f,f_{0})=\frac{1}{2\pi}\int_{0}^{\pi}(f/f_{0}-1)^{2}\,d\lambda
,\qquad
h(f,f_{0})=\frac{1}{2\pi}\int_{0}^{\pi}(f/f_{0}-1)^{2}\frac
{f_{0}}{f}\,d\lambda
\]
for $f\,{=}\,F(d,g)$, $f_{0}\,{=}\,F(d_{0},g_{0})$, $d,d_{0}\,{\in}\,(-1/2,1/2)$,
$g,g_{0}\,{\in}\,\mathcal{G}(m,M)$,
\mbox{$0\,{<}\,m\,{<}\,M$}.

%
\begin{lemma} \label{lemdgehge} For any $\varepsilon>0$,
$
|d-d_{0}|\geq\varepsilon\Rightarrow h(f,f_{0})\geq\frac{1}{\pi
}
(\frac{4M}{m})^{-1/2\varepsilon}$.
\end{lemma}
\begin{pf} Without loss of generality, take $d\geq d_{0}$, then,
since $(x-1)^{2}/x\geq x/2$ for $x\geq4$,
\[
h(f,f_{0}) \geq\frac{m}{4\pi M}\int_{0}^{\pi}\I\bigl\{ \lambda
^{-2(d-d_{0})}\geq4M/m\bigr\} \lambda^{-2(d-d_{0})}\, d\lambda
\geq\frac{1}{\pi}\biggl(\frac{4M}{m}\biggr)^{-1/2\varepsilon}.\quad
\]
\upqed
\end{pf}
%
%
\begin{lemma} \label{lemdgebge}There exists $C>0$ such that, for any
$\varepsilon>0$,
\[
|d-d_{0}|\geq\varepsilon\quad\Rightarrow\quad b(f,f_{0})\geq C^{-1/2\varepsilon}.
\]
\end{lemma}
\begin{pf} If $d\geq d_{0}$, then, since $(x-1)^{2}\geq x^{2}/2$
for $x\geq4$,
\[
b(f,f_{0})
\geq\frac{m^{2}}{4\pi M^{2}}\int_{0}^{\pi}\I\bigl\{ \lambda
^{-2(d-d_{0})}\geq4M/m\bigr\} \lambda^{-4(d-d_{0})}\, d\lambda
\geq\frac{4}{\pi}\biggl(\frac{4M}{m}\biggr)^{-1/2\varepsilon}.
\]
Otherwise, if $d<d_{0}$, one has $(x-1)^{2}\geq1/4$ for $0\leq x\leq1/2$,
so
\[
b(f,f_{0}) \geq\frac{1}{8\pi}\int_{0}^{\pi}\I\bigl\{ \lambda
^{2(d_{0}-d)}\leq m/2M\bigr\} \, d\lambda
\geq\frac{1}{8\pi}\biggl(\frac{2M}{m}\biggr)^{-1/2\varepsilon}.
\]
\upqed
\end{pf}
%
%
\begin{lemma} \label{ineqbh} For any $\tau\in(0,1/4)$, there
exists $C>0$ such that
\[
d-d_{0}<\tfrac{1}{4}-\tau\quad\Rightarrow\quad b(f,f_{0})\leq Ch(f,f_{0}).\vadjust{\goodbreak}
\]
\end{lemma}
\begin{pf} If $d\leq d_{0}$, the bound is trivial, since $f/f_{0}\leq
M/m\pi^{2(d_{0}-d)}$.
Assume $d>d_{0}$, and let $A\geq1/2$ some arbitrary large constant.
Since $(x-1)^{2}\leq x^{2}$ for $x\geq1/2$, one has
%
%
\begin{eqnarray}\label{eqboundb}
b(f,f_{0})
& \leq& Ah(f,f_{0})+\frac{M^{2}}{2\pi m^{2}}\int_{0}^{\pi}\I
\{
f(\lambda)/f_{0}(\lambda)\geq A\} \lambda
^{-4(d-d_{0})}\,d\lambda
\nonumber\\
& \leq& Ah(f,f_{0})+\frac{M^{2}}{2\pi m^{2}}\int_{0}^{\pi}\I
\bigl\{
\la^{-2(d-d_{0})}\geq Am/M\bigr\} \lambda^{-4(d-d_{0})}\,d\lambda
\\
& \leq& Ah(f,f_{0})+\frac{C'(Am/M)^{2-1/2(d-d_{0})}}{1-4t},
\nonumber
\end{eqnarray}
provided $A\geq M/m$ and $C'=M^{2}/2\pi m^{2}$. In turn, since
$(x-1)^{2}\geq x^{2}/2$ for $x\geq4$, and assuming $A\geq4M^{2}/m^{2}$,
then $\la^{-2(d-d_{0})}\geq Am/M$ implies that $f/f_{0}\geq
Am^{2}/M^{2}\geq4$,
and $(f/f_{0}-1)^{2}f_{0}/f\geq f/2f_{0}\geq Am^{2}/2M^{2}$. Therefore
%
%
\begin{eqnarray}
h(f,f_{0}) & \geq& \frac{1}{2\pi}\int_{0}^{\pi}\I\bigl\{ \lambda
^{-2(d-d_{0})}\geq Am/M\bigr\} (f/f_{0}-1)^{2}\frac{f_{0}}{f}\,
d\lambda\\
\label{eqboundh}
& \geq& (Am/M)^{2-1/2(d-d_{0})}/4\pi A.
\end{eqnarray}
One concludes by combining (\ref{eqboundb}) with (\ref{eqboundh})
and taking $A=4M^{2}/m^{2}$.
\end{pf}

The lemma below makes the same assumptions with respect to $f$ and
$f_0$.
%
%
\begin{lemma} \label{lemdahlhaus}
$ d>d_0 \Rightarrow \KL_n(f_0;f) \geq\frac{m^2}{M^2\pi^{2}}
b_n(f_0,f). $
\end{lemma}
\begin{pf}
Dahlhaus [(\citeyear{dhl89}), page 1755] proves that $ \KL_n(f_0;f) \geq C^{-2}
b_n(f_0$, $f)$ where
$C$ is the largest eigenvalue of $T_n(f_0)T_n^{-1}(f)$. In our case,
$f_0/f\leq M\pi^{2(d-d_{0})}/m $, hence $C^{-2}=m^2/M^2\pi^{2(d-d_0)}$.
\end{pf}

The last lemma applies to the FEXP formulation of Section~\ref{secconsfexp}.
%
%
\begin{lemma} \label{lemboundh} For $\varepsilon\in(0,1/4)$,
$f_{0}(\la)=(2-2\cos\la)^{-d_{0}}\exp\{
w_{0}(\la)\}$, $f(\la)=(2-2\cos\la)^{-d}\exp
\{
w(\la)\}, $
one has
\[
|d-d_{0}|\leq\varepsilon,\qquad|w-w_{0}|\leq
\varepsilon
\quad\Rightarrow\quad h(f,f_{0})\leq7\varepsilon.
\]
\end{lemma}
\begin{pf}
Without loss of generality, take $d-d_{0}\geq0$. Then $f_{0}/f-1\leq
2^{\varepsilon}e^{\varepsilon}-1\leq(1+\log2)\varepsilon$,
since $e^{x}\leq1+2x$ for $x\in[0,1]$. Moreover, since $2(1-\cos\la
)\geq
\la^{2}/3$
for $\la\in(0,\pi)$, one has
\[
\int_{0}^{\pi}\frac{f(\lambda)}{f_{0}(\lambda)}\, d\lambda
=e^{\varepsilon}3^{(d-d_{0})}\int_{0}^{\pi}\la^{-2(d-d_{0})}\,
d\lambda
\leq\frac{\pi e^{\varepsilon}3^{\varepsilon}}{1-2\varepsilon}
\]
and, to conclude, as again $e^{x}\leq1+2x$ for $x\in[0,1]$, and
$e^{\varepsilon(1+\log3)}(1-2\varepsilon)^{-1}-1\leq10\varepsilon
$, for
$\varepsilon\leq1/4$,
\[
h(f,f_{0})=\frac{1}{2\pi}\int_{0}^{\pi}\biggl(\frac{f(\lambda
)}{f_{0}(\lambda)}+\frac{f_{0}(\lambda)}{f(\lambda)}-2\biggr)\,
d\lambda
\leq(6+\log2)\varepsilon.
\]
\upqed
\end{pf}
\end{appendix}

\section*{Acknowledgments}

Part of this work was done while the third author was visiting the
Universit\'{e} Paris Dauphine, CEREMADE, whom he thanks for warm
hospitality and financial support.

\begin{supplement}
\stitle{Bayesian nonparametric estimation of the spectral density of a long
or intermediate memory Gaussian process: Supplementary material\\}
\slink[doi]{10.1214/11-AOS955SUPP} 
\sdatatype{.pdf}
\sfilename{aos955\_supp.pdf}
\sdescription{Proof of technical lemmas and theorems stated in the paper.}
\end{supplement}

%

\printaddresses

\end{document}